\documentclass[11pt,letterpaper]{article}
\usepackage{amsmath,pgf,pgfarrows,pgfnodes,float,appendix, hyperref,scalerel,amssymb}
\usepackage{graphicx}
\usepackage{subfigure}
\usepackage[margin=0.9in]{geometry}
\usepackage{pgfplots}
\usepackage{tikz}
\usepackage[normalem]{ulem}
\usepackage{mathrsfs}
\usepackage{slashed}
\usepackage[compat=1.0.0]{tikz-feynman}
\usepackage{wrapfig}
\DeclareMathOperator{\Tr}{tr}
\usetikzlibrary{arrows}
\pgfplotsset{compat=1.15}
\usetikzlibrary{decorations.markings}
\usepackage{braket}
\usepackage{simpler-wick}
\usepackage{mathtools}
\usepackage{rotating}
\usepackage{float}

\newcommand{\arrowIn}{
\tikz \draw[-stealth] (-1pt,0) -- (1pt,0);
}
\newcommand*\circled[1]{\tikz[baseline=(char.base)]{
            \node[shape=circle,draw,inner sep=2pt] (char) {#1};}}

\def\R{{\cal R}}

\count\footins = 1000
\title{Fixed lines in four fermion models in two dimensions}
\author{Sidan A, Tom Banks\\ 
Department of Physics and NHETC\\
Rutgers University, Piscataway, NJ 08854 USA\\
\href{mailto:sa1975@physics.rutgers.edu}{sidan.aa@rutgers.edu}, \href{mailto:tibanks@ucsc.edu}{tibanks@ucsc.edu}}
\date{}

\begin{document}

\renewcommand{\figurename}{Figure}
\renewcommand\appendixpagename{\centering \large APPENDIX. }
\maketitle

\begin{abstract} 

Motivated by conjectures about near-horizon dynamics in quantum gravity, we search for lines of perturbatively accessible fixed points emanating from models of $N$ free fermions.  Through two loops we find a new class of models, apart from the well-known Abelian Thirring models.  Further study is needed to see whether these can lead to true conformal manifolds, or perhaps a new class of large N fixed points.

\end{abstract}

\section{Introduction}

\hspace{6mm}In a recent paper \cite{hilbertbundles}, one of us proposed a general strategy for studying the local physics of models of quantum gravity in given classical space-time backgrounds. Following Jacobson \cite{ted95}, the classical space-time is assumed to reproduce the hydrodynamics of the underlying quantum theory, assigning a local entropy to each causal diamond according to the area law demonstrated in individual works of Bekenstein, Hawking, Jacobson, and Bousso (BHJB) \cite{bhjb}, which is equivalent to the double null projection of Einstein's equations with some conserved stress tensor source. The ansatz of Carlip and Solodukhin \cite{carlip,solo} (C-S), generalized to arbitrary causal diamonds in \cite{BZ}, implies that the modular Hamiltonian of each causal diamond is the Virasoro generator $L_0$ of a $1 + 1$ dimensional conformal field theory (CFT) living on the (past or future)\footnote{The copies of the CFT on the past and future boundary of the diamond are related by the unitary time evolution operator in the diamond.} {\it stretched horizon}.  The CFT has a UV cutoff, related to the smallest size of causal diamond for which the C-S analysis is valid, and a central charge proportional to the area of the holographic screen of the diamond, in Planck units. Cardy's formula then matches the CFT entropy to the BHJB entropy, with the UV cutoff on the Virasoro spectrum taken just at the Cardy matching point.  

In \cite{hilbertbundles} the CFT was taken to be built from free fermion fields $\psi_i$ where the label $i$ runs over the spectrum of the Dirac operator on the holographic screen of the diamond, with a transverse UV cutoff determined by the finite central charge of the CFT.  The motivation for this choice, had its roots in the Holographic Space-time formalism \cite{HST}, which was in turn based on A. Connes \cite{connes} observation that Riemannian geometry is completely determined by the Dirac operator. One can think of the fields $\psi_i (t,z)$ as the expansion coefficients of the solutions of a fluctuating Dirac equation (which represents a fluctuating geometry of the holographic screen) in eigenspinors of the Dirac operator on the background geometry.   The background geometry represents the coarse-grained hydrodynamic average of the system.   

The authors of \cite{lshpss} have argued that the dynamical system on the stretched horizon of a black hole must be a fast scrambler of quantum information. Free fields are not fast scramblers.  Bilinears in spinors on the transverse manifold, can be used to construct p-forms of arbitrary rank.  Since the spinors are also $1+1$ dimensional free fields, two such bilinears can also be commuting $U(1)$ currents, whose rank adds up to the dimension of the transverse holographic screen.  Thus we can construct a conformally invariant Thirring interaction, which is formally invariant under volume preserving maps of the screen. This interaction does not respect distance constraints on the holoscreen and is plausibly a fast scrambler.  The Thirring interaction is a line of fixed points passing through the free field point.

The preservation of the central charge is a property of lines of fixed points in two dimensions.  The purpose of the present paper is to explore whether there can be other lines of fixed points emanating from the free fermion point.  A necessary, but not sufficient condition for this is the vanishing of the one and two-loop beta functions.  We will explore this condition for general four-fermion interactions.

\section{The Generalized Thirring model}\label{sec:thirring}
\hspace{6mm}Before defining the generalized Thirring model, we will introduce the original Thirring model \cite{Thirring:1958in} in two dimensions. The original Thirring model of a massless Dirac fermion with a four-fermion interaction of the current-current type has the following Lagrangian,
\begin{equation}
\begin{split}
   \mathcal{L}_{Th} &= i\overline{\psi}\gamma^{\mu}\partial_{\mu}\psi + \frac{1}{2}gj^{\mu}j_{\mu}  \hspace{1mm}, 
\end{split}
\end{equation}
with $g$ being the Thirring coupling and the current is defined as $j^{\mu} = \overline{\psi}\gamma^{\mu}\psi$. This interaction is unique because any four-fermi interactions can be rewritten in terms of this current-current form using the Fierz transformation, a mathematical procedure used to deal with expressions involving products of Dirac gamma matrices and spinors. A derivation of the Fierz identity can be found in the appendix. \ref{sec:appendixA}. $\overline{\psi}=\psi^{\dagger}\gamma^0$ is the Dirac adjoint of a two-component complex fermion $\psi=\begin{pmatrix} \psi_1 \\
\psi_2
\end{pmatrix}$. The equations of motion for $\overline{\psi}$ and $\psi$ are determined to be
\begin{equation}\label{eq:thirringeom}
    \begin{split}
        i\gamma^{\mu}\partial_{\mu}\psi-gj^{\mu}\gamma_{\mu}\psi&=0 \hspace{1mm},\\
        i\partial_{\mu}\overline{\psi}\gamma^{\mu}+ gj_{\mu}\overline{\psi}\gamma^{\mu}&=0 \hspace{1mm},
    \end{split}
\end{equation}
respectively. Then we find that the conservation law is given by $\partial_{\mu}j^{\mu} = 0$. The theory is renormalizable because the coupling constant is dimensionless. In addition, the Lagrangian has $U(1)$ symmetry $\psi \rightarrow e^{i\alpha} \psi$ and chiral symmetry $\psi \rightarrow e^{i\alpha\gamma^3}\psi$ with parameter $\alpha$. In this paper we choose the Weyl representation where $\gamma^0 = \hspace{1mm}\begin{pmatrix}
    0 & 1\\
    1 & 0
\end{pmatrix}\hspace{1mm}$ $\gamma^1=\begin{pmatrix}
    0 & 1\\
    -1 & 0
\end{pmatrix}$, and $\gamma^3=\gamma^0\gamma^1 = \begin{pmatrix}
    -1 & 0\\
    0 & 1
\end{pmatrix} $ since it is more convenient for massless theories. 

For the generalized Thirring model, we consider an arbitrary number $N$ massless fermions, denote different fermions with upper indices $a, b, \dots = 1, \dots, N$, then the Lagrangian of the theory can written as 
\begin{equation}
    \begin{split}
        \mathcal{L}&= i\overline{\psi}^a \gamma^{\mu}\partial_{\mu}\psi^a + R_{abcd}\overline{\psi}^a\gamma^{\mu}\psi^b\overline{\psi}^c\gamma_{\mu}\psi^d \hspace{1mm},
    \end{split}
\end{equation}
where $R_{abcd}$'s are coupling constants, and repeated indices are summed over $1, \dots, N$ in the expression. In two dimensions, fermionic fields have mass dimension $1/2$, and $R_{abcd}$ is dimensionless, thus the theory is renormalizable. This model has minimum symmetry. Consider global transformations $\psi^a \rightarrow e^{i\lambda_a}\psi^a$ and chiral transformations $\psi^a \rightarrow e^{i\lambda_a \gamma^3}\psi^a$ in each fields with parameter $\lambda_a$. These transformations in general break both the global and chiral symmetry unless specific conditions are imposed. As we can see, for any interaction $R_{abcd}\overline{\psi}^a\gamma^{\mu}\psi^b\overline{\psi}^c\gamma_{\mu}\psi^d$ to be invariant under global and chiral transformations, we need the parameters to satisfy $\lambda_a-\lambda_b+\lambda_c-\lambda_d=0$. A special case of this model is when $a=b$ and $c=d$, this is known as the Abelian Thirring model, which is invariant under the global and chiral transformations. 

Write down the current-current interaction explicitly in terms of spinor components to obtain symmetry properties in the indices of the current-current couplings. These properties come purely from the Grassmannian nature of fermions, meaning that they are self-anticommuting.
\begin{equation}\label{eq:4fermiintminkowski}
    \begin{split}
        \overline{\psi}^a\gamma^{\mu}\psi^b\overline{\psi}^c\gamma_{\mu}\psi^d &= g_{\mu\nu}\overline{\psi}^a\gamma^{\mu}\psi^b\overline{\psi}^c\gamma^{\nu}\psi^d 
        = 2\psi^{a\dagger}_1\psi^b_1\psi^{c\dagger}_2\psi^d_2 + 2\psi^{a\dagger}_2\psi^b_2\psi^{c\dagger}_1\psi^d_1 
    \end{split}
\end{equation}
where the metric $g_{\mu\nu} = diag(+1, -1)$ and the lower indices are spinor indices of the two-dimensional fermionic fields. 

From Eq.(\ref{eq:4fermiintminkowski}) we see that for Dirac fermions, couplings $R_{abcd}$ have the following conditions in their indices:
\begin{equation}
    R_{abcd} = R_{cdab} \hspace{1mm}.
\end{equation}
On the other hand, Majorana fermions are real fields, i.e. $\psi^{\dagger}=\psi$, which give additional symmetry properties in the indices. The Majorana case couplings $R_{abcd}$ are also antisymmetric in their first and the second pair of indices, 
\begin{equation}
    R_{abcd} = -R_{bacd} = -R_{abdc} \hspace{1mm}.
\end{equation}

Now we will use the renormalization method to remove the UV divergence that appeared in the Feynmann diagrams due to unknown small-scale physics, we separate the fields and couplings into a renormalized part and a counterterm, where the counterterm is intrinsically infinite and removes the UV divergence in the diagrams. Let $\psi^a_0$ and $R_{abcd}^0$ be the bare quantities, and let $\psi^a$ and $R_{abcd}$ be renormalized ones. Up to two-loop level, 
\begin{equation}
    \begin{split}
        \psi^a_0 = \sqrt{Z_{a}}\psi^a \hspace{3mm} &\text{with} \hspace{3mm} Z_{a}=1+\delta^{(1)}_{a}+\delta^{(2)}_{a}, \\ R^0_{abcd} = Z_{abcd}R_{abcd} \hspace{3mm} &\text{with} \hspace{3mm} Z_{abcd}=1 + \delta^{(1)}_{abcd} + \delta^{(2)}_{abcd} ,
    \end{split}
\end{equation}
where $Z$'s are the renormalization factors of the fields and coupling constants, $\delta^{(k)}_{a}$ and $\delta^{(k)}_{abcd}$ are the counterterms at k-loop level. Note that the field strength renormalization is $\psi^a_0=\sqrt{Z^a_b}\psi^b$ with $b$ summed over $1,\dots, N$, but as we will show that at one and two-loop levels off-diagonal entries in $Z^a_b$ vanish, therefore, we denote diagonal terms $Z_a\equiv Z^a_a$ for convenience. We also define $\tilde{Z}_{abcd}=(Z_aZ_bZ_cZ_d)^{1/2}Z_{abcd} = 1+\tilde{\delta}^{(1)}_{abcd}+\tilde{\delta}^{(2)}_{abcd}$ for convenience, note that here repeated indices are NOT summed over. Using the above expressions to rewrite the Lagrangian that was originally written in terms of bare quantities, we can separate it into the renormalized and the counterterm parts,
\begin{equation}\label{eq:renormlag}
    \begin{split}
        \mathcal{L} &= \sum_{a=1,\dots,N}i\overline{\psi}^a\gamma^{\mu}\partial_{\mu}\psi^a + \sum_{\substack{a,b,c,d \\ =1,\dots,N}}R_{abcd}\overline{\psi}^a\gamma^{\mu}\psi^b\overline{\psi}^c\gamma_{\mu}\psi^d \\
        &\hspace{15mm} + \sum_{a=1,\dots,N}i\delta^{(1)}_a\overline{\psi}^a\gamma^{\mu}\partial_{\mu}\psi^a + \sum_{\substack{a,b,c,d \\ =1,\dots,N}}\tilde{\delta}^{(1)}_{abcd}R_{abcd}\overline{\psi}^a\gamma^{\mu}\psi^b\overline{\psi}^c\gamma_{\mu}\psi^d \\
        &\hspace{30mm} + \sum_{a=1,\dots,N}i\delta^{(2)}_a\overline{\psi}^a\gamma^{\mu}\partial_{\mu}\psi^a + \sum_{\substack{a,b,c,d \\ =1,\dots,N}}\tilde{\delta}^{(2)}_{abcd}R_{abcd}\overline{\psi}^a\gamma^{\mu}\psi^b\overline{\psi}^c\gamma_{\mu}\psi^d .
    \end{split}
\end{equation}

The idea of renormalization is that, at $L$-loop level, the $L$-loop counterterm absorbs the divergence from $L$-loop diagrams, thus making the correlation function correction finite at this level. To compute the $\beta$-function of $R_{abcd}$, we need to determine the counterterms. Treat the last three terms in Lagrangian Eq.(\ref{eq:renormlag}) as a perturbation and the correlation function is given by
\begin{equation}\label{eq:correlationfcn}
    \begin{split}
        \braket{\Omega|\mathcal{T}\{\overline{\psi}^a\gamma^{\mu}\psi^b\overline{\psi}^c\gamma_{\mu}\psi^d\}|\Omega} &= \lim_{T\rightarrow \infty(1-i\epsilon)}\frac{\braket{0|\mathcal{T}\{\overline{\psi}^a\gamma^{\mu}\psi^b\overline{\psi}^c\gamma_{\mu}\psi^d \exp[-i\int_{-T}^T dt H_I(t)] \}|0}}{\braket{0|\mathcal{T}\{\exp[-i\int_{-T}^T dt H_I(t)]\} |0}} \\
        &= \text{Sum of all connected and amputated diagrams} \\
        & \hspace{5mm} \text{(By Wick's theorem) }
    \end{split}
\end{equation}

Besides standard Feynman rules for fermions, we define a new rule in momentum space for the interaction vertices $\overline{\psi}^a\gamma^{\mu}\psi^b\overline{\psi}^c\gamma_{\mu}\psi^d$, and the counterterms $i\delta_a\overline{\psi}^a\gamma^{\mu}\partial_{\mu}\psi^a, \tilde{\delta}_{abcd}R_{abcd}\overline{\psi}^a\gamma^{\mu}\psi^b\overline{\psi}^c\gamma_{\mu}\psi^d$ as shown in Figure.\ref{fig:feynmanrules}.

\begin{figure}[H]
    \begin{minipage}{0.34\textwidth}
    \begin{tikzpicture}[scale=0.4]
    \centering
        \draw[black] (0.1,0)--(1,1) node[sloped, pos=0.5, allow upside down]{\arrowIn};;
        \draw[black] (-0.1,0)--(-1,1) node[sloped, pos=0.5, allow upside down]{\arrowIn};;
        \draw[black] (1,-1)--(0.1,0) node[sloped, pos=0.5, allow upside down]{\arrowIn};;
        \draw[black] (-1,-1)--(-0.1,0) node[sloped, pos=0.5, allow upside down]{\arrowIn};;
        \filldraw (-1,-1) circle (0pt) node[anchor=east]{b};
        \filldraw (1,-1) circle (0pt) node[anchor=west]{d};
        \filldraw (1,1) circle (0pt) node[anchor=west]{c};
        \filldraw (-1,1) circle (0pt) node[anchor=east]{a};
        \filldraw (1.3,0) circle (0pt) node[anchor=west]{$= -iR_{abcd}(\gamma^{\mu}\dots\gamma_{\mu})$};
        \end{tikzpicture}
    \end{minipage}
    \begin{minipage}{0.3\textwidth}
    \begin{tikzpicture}[scale=0.4]
    \centering
        \draw[black] (-1.5,0)--(-0.3,0) node[sloped, pos=0.5, allow upside down]{\arrowIn};;
        \draw[black] (0.3,0)--(1.5,0) node[sloped, pos=0.5, allow upside down]{\arrowIn};;
        \draw[black] (0,0) circle (0.3cm);
        \draw[black] (-0.21,-0.21)--(0.21,0.21);
        \draw[black] (-0.21,0.21)--(0.21,-0.21);
        \filldraw (-1,0.2) circle (0pt) node[anchor=south]{p};
        \filldraw (1,0.2) circle (0pt) node[anchor=south]{p};
        \filldraw (-1.6,0) circle (0pt) node[anchor=east]{b};
        \filldraw (1.6,0) circle (0pt) node[anchor=west]{a};
        \filldraw (2.5,0) circle (0pt) node[anchor=west]{$= -i\slashed{p}\delta_{ab}\delta^{(k)}_a$};
        \end{tikzpicture}
    \end{minipage}
    \begin{minipage}{0.3\textwidth}
    \begin{tikzpicture}[scale=0.4]
    \centering
        \draw[black] (-1,-1)--(0,0) node[sloped, pos=0.5, allow upside down]{\arrowIn};;
        \draw[black] (0.2,0.2)--(1,1) node[sloped, pos=0.5, allow upside down]{\arrowIn};;
        \draw[black] (-0.2,0.2)--(-1,1) node[sloped, pos=0.5, allow upside down]{\arrowIn};;
        \draw[black] (1,-1)--(0,0) node[sloped, pos=0.5, allow upside down]{\arrowIn};;
        \draw[black] (0,0) circle (0.3cm);
        \draw[black] (-0.21,-0.21)--(0.21,0.21);
        \draw[black] (-0.21,0.21)--(0.21,-0.21);
        \filldraw (-1,-1) circle (0pt) node[anchor=east]{b};
        \filldraw (1,-1) circle (0pt) node[anchor=west]{d};
        \filldraw (1,1) circle (0pt) node[anchor=west]{c};
        \filldraw (-1,1) circle (0pt) node[anchor=east]{a};
        \filldraw (1.3,0) circle (0pt) node[anchor=west]{$= -i\tilde{\delta}^{(k)}_{abcd}R_{abcd}(\gamma^{\mu}\dots\gamma_{\mu})$};
        \end{tikzpicture}
    \end{minipage}
    \caption{Feynman diagrams in momentum space for the three interactions in Lagrangian Eq.(\ref{eq:renormlag}). Note that $\delta_{ab}$ is the Kronecker delta, $\delta^{(k)}_a$ is the field strength counter term. In the current-current vertex, if two fermions $\psi^a$ and $\psi^b$ form a current $\overline{\psi}^a\gamma^{\mu}\psi^b$, then their legs join at the vertex. If they do not join, e.g. $\psi^a$ and $\psi^d$, then they are not in the same current.}
    \label{fig:feynmanrules}
\end{figure}

\subsection{Field Strength and Coupling Constant Corrections at the One-loop Level}\label{subsec:1loop}

\hspace{6mm}To determine the field strength counterterm $\delta^{(1)}_{a}$ for the fields, we need to compute the two-point function. For any $a\in \{1, \dots, N\}$, the two-point correlation functions have the following corrections at the one-loop level:
\begin{figure}[H]
    \centering
    \begin{tikzpicture}[scale=0.6]
        \draw[black] (-1.5+1,0)--(-0.2+1,0) node[sloped, pos=0.5, allow upside down]{\arrowIn};;
        \draw[black] (0.2+1,0)--(1.5+1,0) node[sloped, pos=0.5, allow upside down]{\arrowIn};;
        \filldraw (-1+0.8,0.2) circle (0pt) node[anchor=south]{p};
        \filldraw (1+1.3,0.2) circle (0pt) node[anchor=south]{p};
        \filldraw (1,1.2) circle (0pt) node[anchor=south]{q};
        \draw[black] (0.8,0) arc(250:-70:0.6)node[sloped, pos=0.5, allow upside down]{\arrowIn};;
        \filldraw (-6,0.5) circle (0pt) node[anchor=west]{$G_1^{(2)}(x,y)$};
        \filldraw (-2.5,0.5) circle (0pt) node[anchor=west]{$=$};
        \filldraw (2+1.5,0.5) circle (0pt) node[anchor=west]{$+$};
        \draw[black] (-1.5+7,0)--(7,0) node[sloped, pos=0.5, allow upside down]{\arrowIn};;
        \draw[black] (7,0)--(1.5+7,0) node[sloped, pos=0.5, allow upside down]{\arrowIn};;
        \filldraw (-1+6.7,0.2) circle (0pt) node[anchor=south]{p};
        \filldraw (1+7.3,0.2) circle (0pt) node[anchor=south]{p};
        \draw[black] (6.5,0.5) arc(225:-45:0.6)node[sloped, pos=0.5, allow upside down]{\arrowIn};;
        \draw[black] (6.5,0.5)--(6.93,0.2)--(7.35,0.5);
        \filldraw (7,1.5) circle (0pt) node[anchor=south]{q};
        \filldraw (9.5,0.5) circle (0pt) node[anchor=west]{$+$};
        \draw[black] (11.5,0)--(12.7,0) node[sloped, pos=0.5, allow upside down]{\arrowIn};;
        \draw[black] (13.3,0)--(14.5,0) node[sloped, pos=0.5, allow upside down]{\arrowIn};;
        \draw[black] (13,0) circle (0.3cm);
        \draw[black] (-0.21+13,-0.21)--(0.21+13,0.21);
        \draw[black] (-0.21+13,0.21)--(0.21+13,-0.21);
        \filldraw (-1+13,0.2) circle (0pt) node[anchor=south]{p};
        \filldraw (14,0.2) circle (0pt) node[anchor=south]{p};
        \end{tikzpicture}
    \caption{Feynman diagrams at one-loop level for the two-point function.}
    \label{fig:2pt1loop}
\end{figure}

Write down the momentum space correlation function at the one-loop level for each diagram, the first diagram in Figure.\ref{fig:2pt1loop} contributes
\begin{equation}\label{eq:2pt1loopdiag1}
    \begin{split}
        -iR_{efgh}\braket{ \wick{\Omega|
        \c3{\overline{\psi}^a_x} (\c4{\overline{\psi}^e_z} \gamma^{\mu}\c2\psi^f_z \c2{\overline{\psi}^g_z} \gamma_{\mu} \c3\psi^h_z) \c4{\psi^b_y}|\Omega}} &\Rightarrow \delta_{ah}\delta_{be}\delta_{fg}\frac{-iR_{efgh}}{2}\int \frac{d^2q}{(2\pi)^2}\frac{i\slashed{p}}{p^2}\gamma^{\mu}\frac{i\slashed{q}}{q^2+i\epsilon}\gamma_{\mu}\frac{i\slashed{p}}{p^2} \\
        &= -\frac{R_{beea}}{2}\frac{p_{\alpha}p_{\beta}}{p^4}\gamma^{\alpha}\gamma^{\mu}\gamma^{\nu}\gamma_{\mu}\gamma^{\beta}\int \frac{d^2q}{(2\pi)^2}\frac{q_{\nu}}{q^2+i\epsilon} = 0,
    \end{split}
\end{equation}
where index $e$ is summed over $1, \dots, N$. The expression vanishes because both $\gamma^{\mu}\gamma^{\nu}\gamma_{\mu}=0$ by gamma matrices identities in Eq.(\ref{eq:gammaidentities}) and the integrand is an odd function of $q$ and it is integrated over $(-\infty, \infty)$ thus gives 0. This calculation indicates that any diagram that has one-end-open loops has 0 amplitude since $\gamma^{\mu}(\cdots)\gamma_{\mu}$ where $(\cdots)$ consists of an odd number of Dirac matrices vanishes.

Similarly, the second diagram in Figure.\ref{fig:2pt1loop} contributes
\begin{equation}\label{eq:2pt1loopdiag2}
    \begin{split}
        -iR_{efgh}\braket{ \wick{\Omega|
        \c2{\overline{\psi}^a_x} (\c3{\overline{\psi}^e_z} \gamma^{\mu}\c2 \psi^f_z \c2{\overline{\psi}^g_z} \gamma_{\mu} \c2\psi^h_z) \c3{\psi^b_y}|\Omega}} &\Rightarrow \delta_{be}\delta_{af}\delta_{gh}\frac{iR_{efgh}}{2} \frac{i\slashed{p}}{p^2}\gamma^{\mu}\frac{i\slashed{p}}{p^2} \int \frac{d^2q}{(2\pi)^2}\Tr\bigg(\frac{i\slashed{q}}{q^2+i\epsilon}\gamma_{\mu}\bigg) \\
        &= \frac{R_{baee}}{2} \frac{p_{\alpha}p_{\beta}}{p^4}\gamma^{\alpha}\gamma^{\mu}\gamma^{\beta} \int \frac{d^2q}{(2\pi)^2}\frac{q_{\mu}}{q^2+i\epsilon} =0,
    \end{split}
\end{equation}
with index $e$ summed over $1, \dots, N$. This diagram also vanishes since the integrand is an odd function in $q$ integrated over $q\in(-\infty, \infty)$. Since the sum of the three diagrams is zero, we conclude that, at the one-loop level, the field-strength correction $\delta^{(1)}_{a}=0$.

Now let us compute the current-current correlation function to determine the one-loop counterterm to the coupling $R_{abcd}$. Recall that for the $\phi^4$ theory at the one-loop level, we have three diagrams, i.e. the s, t, and u-channels, contributing corrections to the $\phi^4$ coupling constant $\lambda$. In analogy to the scalar theory, we have diagrams in the three channels, but the current-current vertex in our theory generates multiple diagrams in each channel due to the broken permutation symmetry of this vertex. However, the observation we made from the calculations above tells us that, using $\gamma$-matrices identities in two dimensions and features of integration over undetermined momenta, we can eliminate most of the diagrams. The trick is presented in appendix.\ref{sec:appendixB}.

At the one-loop level, the $288$ Wick contractions reduce to $10$ distinct Feynman diagrams, using the trick explained in appendix.\ref{sec:appendixB}, $8$ more vanish and there are only $2$ diagrams left that contribute to the correlation function. For any choice of $a, b, c, d \in \{1, \dots, N\}$, the correlation function of the current-current interaction is given by
\begin{figure}[H]
    \centering
    \begin{tikzpicture}[scale=0.6]
        \draw[black] (2,-0.1) arc(-35:-145:1.23)node[sloped, pos=0.5, allow upside down]{\arrowIn};;
        \draw[black] (2,0.1) arc(35:145:1.23)node[sloped, pos=0.5, allow upside down]{\arrowIn};;
        \draw[black] (1,2)--(2,0.1) node[sloped, pos=0.5, allow upside down]{\arrowIn};;
        \draw[black] (0,0.1)--(1,2) node[sloped, pos=0.5, allow upside down]{\arrowIn};;
        \draw[black] (1,-2)--(2,-0.1) node[sloped, pos=0.5, allow upside down]{\arrowIn};;
        \draw[black] (0,-0.1)--(1,-2) node[sloped, pos=0.5, allow upside down]{\arrowIn};;
        \filldraw (0.5,-1.5) circle (0pt) node[anchor=east]{c};
        \filldraw (1.45,-1.43) circle (0pt) node[anchor=west]{d};
        \filldraw (0.5,1.5) circle (0pt) node[anchor=east]{a};
        \filldraw (1.45,1.57) circle (0pt) node[anchor=west]{b};
        \filldraw (-6,0) circle (0pt) node[anchor=west]{$G_1^{(4)}(x,y)$};
        \filldraw (-2.5,0) circle (0pt) node[anchor=west]{$=$};
        \filldraw (2+1.5,0) circle (0pt) node[anchor=west]{$+$};
        \draw[black] (6+2,-0.1) arc(-35:-145:1.23)node[sloped, pos=0.5, allow upside down]{\arrowIn};;
        \draw[black] (4+2,0.1) arc(145:35:1.22)node[sloped, pos=0.5, allow upside down]{\arrowIn};;
        \draw[black] (6+2,0.1) arc(35:56:4);
        \draw[black] (4+2,0.1) arc(145:127:4);
        \draw[black] (5+2,1.09) arc(239:120:0.56)node[sloped, pos=0.5, allow upside down]{\arrowIn};;
        \draw[black] (5+2,2.05) arc(65:-52:0.5)node[sloped, pos=0.5, allow upside down]{\arrowIn};;
        \draw[black] (5+2,-2)--(6+2,-0.1) node[sloped, pos=0.5, allow upside down]{\arrowIn};;
        \draw[black] (4+2,-0.1)--(5+2,-2) node[sloped, pos=0.5, allow upside down]{\arrowIn};;
        \filldraw (4.5+2,-1.5) circle (0pt) node[anchor=east]{c};
        \filldraw (5.45+2,-1.43) circle (0pt) node[anchor=west]{d};
        \filldraw (4.5+2,1.5) circle (0pt) node[anchor=east]{a};
        \filldraw (5.45+2,1.57) circle (0pt) node[anchor=west]{b};
        \filldraw (9.5,0) circle (0pt) node[anchor=west]{$+$};
        \draw[black] (4.8+8,0.2) arc(225:120:1.05)node[sloped, pos=0.5, allow upside down]{\arrowIn};;
        \draw[black] (5+8,1.85) arc(65:-51:1)node[sloped, pos=0.5, allow upside down]{\arrowIn};;
        \draw[black] (13,0) circle (0.3cm);
        \draw[black] (-0.21+13,-0.21)--(0.21+13,0.21);
        \draw[black] (-0.21+13,0.21)--(0.21+13,-0.21);
        \draw[black] (13.05,-1.85) arc(-60:48:1.03)node[sloped, pos=0.5, allow upside down]{\arrowIn};;
        \draw[black] (13-0.2,-0.2) arc(130:245:1)node[sloped, pos=0.5, allow upside down]{\arrowIn};;
        \filldraw (4.5+8,-1.5) circle (0pt) node[anchor=east]{c};
        \filldraw (5.45+8,-1.43) circle (0pt) node[anchor=west]{d};
        \filldraw (4.5+8,1.5) circle (0pt) node[anchor=east]{a};
        \filldraw (5.45+8,1.57) circle (0pt) node[anchor=west]{b};
        \end{tikzpicture}
    \caption{Feynman diagrams at one-loop level for the four-point function.}
    \label{fig:4pt1loop}
\end{figure}

The first diagram in Figure.\ref{fig:4pt1loop} contributes
\begin{equation}
    \begin{split}
        & (-iR_{e_1f_1g_1h_1})(-iR_{e_2f_2g_2h_2})\braket{\Omega | \wick{\c3{\overline{\psi}^a_x} \gamma^{\mu}\c2\psi^b_x (\c2{\overline{\psi}^{e_1}_z} \gamma^{\nu}\c2\psi^{f_1}_z\c5{\overline{\psi}^{g_1}_z}\gamma_{\nu}\c4\psi^{h_1}_z) (\c2{\overline{\psi}^{e_2}_w }\gamma^{\rho}\c3\psi^{f_2}_w\c4{\overline{\psi}^{g_2}_w}\gamma_{\rho}\c4\psi^{h_2}_w)\c4{\overline{\psi}^c_y}\gamma_{\mu}\c5\psi^d_y}|\Omega} \\
        \Rightarrow  & -R_{eafc}R_{bedf}\int \frac{d^2q}{(2\pi)^2}\Tr\bigg(\frac{i\slashed{p}_b}{p_b^2}\gamma^{\nu}\frac{i\slashed{q}}{q^2+i\epsilon}\gamma^{\rho}\frac{i\slashed{p}_a}{p_a^2}\gamma^{\mu}\bigg)\Tr\bigg(\frac{i\slashed{p}_d}{p_d^2}\gamma_{\nu}\frac{i(\slashed{p}_a-\slashed{p}_c-\slashed{q})}{(p_a-p_c-q)^2+i\epsilon}\gamma_{\rho}\frac{i\slashed{p}_c}{p_c^2}\gamma_{\mu}\bigg) \\
        =& R_{eafc}R_{bedf}P\Gamma_1 \int \frac{d^2q}{(2\pi)^2}\frac{q_{\alpha}(p_a-p_c-q)^{\beta}}{(q^2+i\epsilon)[(p_a-p_c-q)^2+i\epsilon]}, \\
        & \text{where} \hspace{3mm} \Gamma_1 = \Tr(\gamma^b\gamma^{\nu}\gamma^{\alpha}\gamma^{\rho}\gamma^a\gamma^{\mu})\Tr(\gamma_d\gamma_{\nu}\gamma_{\beta}\gamma_{\rho}\gamma_c\gamma_{\mu}) \hspace{3mm} \text{and} \hspace{3mm} P=\frac{(p_a)_a(p_b)_b(p_c)^c(p_d)^d}{p_a^2p_b^2p_c^2p_d^2}.
    \end{split}
\end{equation}

The second diagram contributes
\begin{equation}
    \begin{split}
        & (-iR_{e_1f_1g_1h_1})(-iR_{e_2f_2g_2h_2})\braket{\Omega | \wick{\c3{\overline{\psi}^a_x} \gamma^{\mu}\c2\psi^b_x (\c2{\overline{\psi}^{e_1}_z} \gamma^{\nu}\c2\psi^{f_1}_z\c4{\overline{\psi}^{g_1}_z}\gamma_{\nu}\c5\psi^{h_1}_z) (\c2{\overline{\psi}^{e_2}_w }\gamma^{\rho}\c3\psi^{f_2}_w\c3{\overline{\psi}^{g_2}_w}\gamma_{\rho}\c4\psi^{h_2}_w)\c5{\overline{\psi}^c_y}\gamma_{\mu}\c3\psi^d_y}|\Omega} \\
        \Rightarrow  & -R_{eadf}R_{befc}\int \frac{d^2q}{(2\pi)^2}\Tr\bigg(\frac{i\slashed{p}_b}{p_b^2}\gamma^{\nu}\frac{i\slashed{q}}{q^2+i\epsilon}\gamma^{\rho}\frac{i\slashed{p}_a}{p_a^2}\gamma^{\mu}\bigg)\Tr\bigg(\frac{i\slashed{p}_d}{p_d^2}\gamma_{\rho}\frac{i(\slashed{p}_b-\slashed{p}_c+\slashed{q})}{(p_b-p_c+q)^2+i\epsilon}\gamma_{\nu}\frac{i\slashed{p}_c}{p_c^2}\gamma_{\mu}\bigg) \\
        =& R_{eadf}R_{befc}P\Gamma_2 \int \frac{d^2q}{(2\pi)^2}\frac{q_{\alpha}(p_b-p_c+q)^{\beta}}{(q^2+i\epsilon)[(p_b-p_c+q)^2+i\epsilon]}, \\
        & \text{where} \hspace{3mm} \Gamma_2 = \Tr(\gamma^b\gamma^{\nu}\gamma^{\alpha}\gamma^{\rho}\gamma^a\gamma^{\mu})\Tr(\gamma_d\gamma_{\rho}\gamma_{\beta}\gamma_{\nu}\gamma_c\gamma_{\mu}) = \Gamma_1 \hspace{3mm} \text{since} \hspace{3mm} \gamma_{\rho}\gamma_{\beta}\gamma_{\nu} = \gamma_{\nu}\gamma_{\beta}\gamma_{\rho}.
    \end{split}
\end{equation}
Note that in the above two diagrams, the indices $e, f$ are summed over $1, \dots, N$.

The counterterm contributes in two ways
\begin{equation}
    \begin{split}
        & -i\tilde{\delta}^{(1)}_{efgh}R_{efgh}\braket{ \Omega|\wick{\c3{\overline{\psi}^a_x} \gamma^{\mu}\c2\psi^b_x (\c2{\overline{\psi}^e_z} \gamma^{\nu}\c3\psi^f_z\c3{\overline{\psi}^g_z}\gamma_{\nu}\c2\psi^h_z) \c2{\overline{\psi}^c_y}\gamma_{\mu}\c3\psi^d_y}|\Omega} \Rightarrow -i\tilde{\delta}^{(1)}_{badc}R_{badc}P\Gamma, \\
        &\text{where} \hspace{3mm} \Gamma = \Tr(\gamma^b\gamma^{\nu}\gamma^a\gamma^{\mu})\Tr(\gamma_d\gamma_{\nu}\gamma_c\gamma_{\mu}). \\
        & -i\tilde{\delta}^{(1)}_{efgh}R_{efgh}\braket{ \Omega|\wick{\c3{\overline{\psi}^a_x} \gamma^{\mu}\c2\psi^b_x (\c2{\overline{\psi}^e_z} \gamma^{\nu}\c4\psi^f_z\c2{\overline{\psi}^g_z}\gamma_{\nu}\c3\psi^h_z) \c4{\overline{\psi}^c_y}\gamma_{\mu}\c2\psi^d_y}|\Omega} \Rightarrow i\tilde{\delta}^{(1)}_{dabc}R_{dabc}P\Gamma', \\
        &\text{where} \hspace{3mm} \Gamma' = \Tr(\gamma^b\gamma^{\nu}\gamma_c\gamma_{\mu}\gamma_d\gamma_{\nu}\gamma^a\gamma^{\mu}) = 0 \hspace{3mm} \text{thus this term does not contribute}.
    \end{split}
\end{equation}

To evaluate the integral over undetermined momentum, we apply Feynman parametrization with $A=q^2+i\epsilon$ and $B=(p_a-p_c-q)^2+i\epsilon$,
\begin{equation}
   \frac{1}{AB} = \int_0^1 dx\frac{1}{[xA+(1-x)B]^2} = \int_0^1 dx\frac{1}{(l^2-\Delta)^2},
\end{equation}
where $l\equiv q-(p_a-p_c)(1-x)$ and $\Delta \equiv (p_a-p_c)^2(x^2-x) - i\epsilon$
and the integral over undetermined momentum gives
\begin{equation}\label{eq:feynmanpara}
\begin{split}
    \int \frac{d^2q}{(2\pi)^2}\frac{q_{\alpha}(p_a-p_c-q)^{\beta}}{(q^2+i\epsilon)[(p_a-p_c-q)^2+i\epsilon]} &=  \int_0^1dx\int \frac{d^2l}{(2\pi)^2}\frac{-l_{\alpha}l^{\beta}+B^{\beta}l_{\alpha}+C}{(l^2-\Delta)^2}. 
\end{split}    
\end{equation}
We use $l_{\mu}l_{\nu}=\frac{1}{2}g_{\mu\nu}l^2$ which gives $A=-1/2$. $B, C$ are functions of $p_a-p_c$, and we can ignore $B$ since $l_{\alpha}$ is an odd function which gives 0 when integrated over $l\in [-\infty, \infty]$. 

Now apply dimensional regularization to find
$\int d^2l \frac{l^2}{(l^2-\Delta)^2}$ and $\int d^2l \frac{1}{(l^2-\Delta)^2}$. We have the following results in Minkowski signature
\begin{equation}\label{eq:int1}
    \int \frac{d^dl}{(2\pi)^d} \frac{1}{(l^2-\Delta)^2} = \frac{i}{(4\pi)^{d/2}}\frac{\Gamma(2-d/2)}{\Gamma(2)}\frac{1}{\Delta}^{2-d/2} ,
\end{equation}
\begin{equation}\label{eq:int2}
    \int \frac{d^dl}{(2\pi)^d} \frac{l^{\mu}l^{\nu}}{(l^2-\Delta)^2} = \frac{-i}{(4\pi)^{d/2}}\frac{g^{\mu\nu}}{2}\frac{\Gamma(1-d/2)}{\Gamma(2)}\frac{1}{\Delta}^{1-d/2}
\end{equation}.
The integral in Eq.(\ref{eq:int1}) converges thus compute directly with $d=2$ by wick-rotating to Euclidean space with $l^0=il^0_E$ and $l^2=-l^2_E$. The integral in Eq.(\ref{eq:int2}) diverges logarithmically, we evaluate using dimensional regularization,
\begin{equation} 
\begin{split}
    \int \frac{d^2l}{(2\pi)^2} \frac{1}{(l^2-\Delta)^2} = \int \frac{d^2l_E}{(2\pi)^2} \frac{i}{(l_E^2+\Delta)^2} &= \frac{i}{4\pi \Delta},
\end{split}
\end{equation}
\begin{equation}
\begin{split}
    \int \frac{d^{2-\epsilon}l}{(2\pi)^{2-\epsilon}} \frac{l_{\alpha}l^{\beta}}{(l^2-\Delta)^2} &= \frac{-i{g_{\alpha}}^{\beta}}{8\pi}\frac{\Gamma(\epsilon/2)}{\Gamma(2)}\bigg(\frac{4\pi}{\Delta}\bigg)^{\epsilon/2} \\
    &\rightarrow \frac{-i{g_{\alpha}}^{\beta}}{8\pi}\bigg(\frac{2}{\epsilon} - \# + \mathcal{O}(\epsilon)\bigg)\bigg(1+\frac{\epsilon}{2}\ln\frac{4\pi}{\Delta}+\mathcal{O}(\epsilon^2)\bigg) \sim -\frac{i{g_{\alpha}}^{\beta}}{4\pi\epsilon}, 
\end{split}
\end{equation}
where $\Delta \sim \mu^2$ with $\mu$ as the energy scale. Let $d=2-\epsilon$ and take $\epsilon \rightarrow 0$, note that here $\epsilon$ is the small parameter in dimensional regularization, which is different from the $\epsilon$ in the propagator. Expand $\Gamma(x)$ near $x=0$ we have $\Gamma(x) = 1/x-\#+\mathcal{O}(x)$ where $\#$ is the Euler-Mascheroni constant, and expand $(1/\Delta)^{1-d/2} = 1-(1-d/2)\ln \Delta + \dots$ near $d=2$. We see that the current-current correlation function has a divergence of $1/\epsilon$.

Using $\gamma$ matrices identities to simplify ${g_{\alpha}}^{\beta}\Gamma_1$ where ${g_{\alpha}}^{\beta}$ comes from doing Feynman parametrization, we get
\begin{equation}
    \begin{split}
        {g_{\alpha}}^{\beta}\Gamma_1 &= \Tr(\gamma^b\gamma^{\nu}\gamma^{\alpha}\gamma^{\rho}\gamma^a\gamma^{\mu})\Tr(\gamma_d\gamma_{\nu}\gamma_{\alpha}\gamma_{\rho}\gamma_c\gamma_{\mu}) \\
        &= \Tr(\gamma^b(g^{\nu\alpha}\gamma^{\rho}+g^{\alpha\rho}\gamma^{\nu}-g^{\nu\rho}\gamma^{\alpha})\gamma^a\gamma^{\mu})\Tr(\gamma_d\gamma_{\nu}\gamma_{\alpha}\gamma_{\rho}\gamma_c\gamma_{\mu}) \\
        &= 4\Tr(\gamma^b\gamma^{\nu}\gamma^a\gamma^{\mu})\Tr(\gamma_d\gamma_{\nu}\gamma_c\gamma_{\mu}) = 4\Gamma.
    \end{split}
\end{equation}

Include the symmetry factor in each diagram and the three diagrams in Figure.\ref{fig:4pt1loop} contributes
\begin{equation}
    \begin{split}
        \circled{1} &= \frac{iP\Gamma}{4\pi\epsilon}\sum_{e, f=1}^N R_{eafc}R_{bedf}, \\
        \circled{2} &= -\frac{iP\Gamma}{4\pi\epsilon}\sum_{e, f=1}^N R_{eadf}R_{befc}, \\
        \circled{3} &= -\frac{iP\Gamma}{2}\tilde{\delta}^{(1)}_{badc}R_{badc},
    \end{split}
\end{equation}
and sum up to 0. Up to one-loop level, $\tilde{Z}_{abcd}=(Z_aZ_bZ_cZ_d)^{1/2}\Big(1+\delta^{(1)}_{abcd}\Big)$ and $Z_a = 1 + \delta^{(1)}_a = 1$ because $\delta^{(1)}_a$ is determined to be 0. Then $\Tilde{Z}_{abcd} = 1+\Tilde{\delta}^{(1)}_{abcd}= 1+ \delta^{(1)}_{abcd}$ then we have $\tilde{\delta}^{(1)}_{abcd}=\delta^{(1)}_{abcd}$. Since for Dirac fermions, we do not have antisymmetry in the first and second pair of indices, $R_{abcd}\neq 0$. Then we get
\begin{equation}\label{eq:4pt1loop}
    \begin{split}
        \delta^{(1)}_{abcd}R_{abcd} &= \frac{1}{2\pi\epsilon} \sum_{e, f=1}^N(R_{aecf}R_{ebfd} - R_{aefd}R_{ebcf}) \equiv \frac{\mathcal{R}^{(1)}_{abcd}}{2\pi\epsilon}.
    \end{split}
\end{equation}

\subsection{At Two-loop Level}\label{subsec:2loop}

\hspace{6mm}The correction to the $\beta$-function at the two-loop level is given by
\begin{figure}[H]
    \centering
    \begin{tikzpicture}[scale=0.5]
        \filldraw (-17,0.5) circle (0pt) node[anchor=west]{$G_2^{(2)}(x,y)$};
        \filldraw (-13.3,0.5) circle (0pt) node[anchor=west]{$=$};
        \draw[black] (-1.85-9.75,0.5)--(-0.55-9.55,0.5) node[sloped, pos=0.5, allow upside down]{\arrowIn};;
        \draw[black] (-8.5,0.5)--(1.5-8.5,0.5) node[sloped, pos=0.5, allow upside down]{\arrowIn};;
        \draw[black] (-8.7,0.5)--(-9.9,0.5) node[sloped, pos=0.5, allow upside down]{\arrowIn};;
        \draw[black] (-8.5-1.4,0.5) arc(-180:0:0.7)node[sloped, pos=0.5, allow upside down]{\arrowIn};;
        \filldraw (-11,0.7) circle (0pt) node[anchor=south]{$p$};
        \filldraw (-9.5+1.9,0.7) circle (0pt) node[anchor=south]{$p$};
        \filldraw (-9.3,0.35) circle (0pt) node[anchor=south]{$q_1$};
        \filldraw (-9.2,-0.15) circle (0pt) node[anchor=north]{$q_2$};
        \filldraw (-9.4,1.25) circle (0pt) node[anchor=south]{$q'$};
        \draw[black] (-10.1,0.5) arc(180:0:0.7)node[sloped, pos=0.5, allow upside down]{\arrowIn};;
        \filldraw (-6.4,0.5) circle (0pt) node[anchor=west]{$+$};
        
        \draw[black] (-1.85-3.05,0.5)--(-0.55-3.05,0.5) node[sloped, pos=0.5, allow upside down]{\arrowIn};;
        \draw[black] (-2,0.5)--(1.5-2,0.5) node[sloped, pos=0.5, allow upside down]{\arrowIn};;
        \draw[black] (-3.4,0.5)--(-2.2,0.5) node[sloped, pos=0.5, allow upside down]{\arrowIn};;
        \draw[black] (-2.2,0.5) arc(0:-180:0.6)node[sloped, pos=0.5, allow upside down]{\arrowIn};;
        \filldraw (-5+0.5,0.7) circle (0pt) node[anchor=south]{$p$};
        \filldraw (-3+1.9,0.7) circle (0pt) node[anchor=south]{$p$};
        \filldraw (-2.8,0.4) circle (0pt) node[anchor=south]{$q_1$};
        \filldraw (-2.8,-0.1) circle (0pt) node[anchor=north]{$q_2$};
        \filldraw (-2.8,1.3) circle (0pt) node[anchor=south]{$q'$};
        \draw[black] (-3.6,0.5) arc(180:0:0.8)node[sloped, pos=0.5, allow upside down]{\arrowIn};;
        \filldraw (0,0.5) circle (0pt) node[anchor=west]{$+$};
        
        \draw[black] (-1.5+3,0)--(-0.2+3,0) node[sloped, pos=0.5, allow upside down]{\arrowIn};;
        \draw[black] (0.2+3,0)--(1.5+3,0) node[sloped, pos=0.5, allow upside down]{\arrowIn};;
        \draw[black] (3,0) circle (0.2cm);
        \draw[black] (-0.15+3,-0.15)--(0.15+3,0.15);
        \draw[black] (-0.15+3,0.15)--(0.15+3,-0.15);
        \filldraw (1+0.8,0.2) circle (0pt) node[anchor=south]{$p$};
        \filldraw (3+1.3,0.2) circle (0pt) node[anchor=south]{$p$};
        \filldraw (3,1.2) circle (0pt) node[anchor=south]{$q$};
        \draw[black] (2.8,0) arc(250:-70:0.6)node[sloped, pos=0.5, allow upside down]{\arrowIn};;
        \filldraw (5.1,0.5) circle (0pt) node[anchor=west]{$+$};
        \draw[black] (-1.5+8,0)--(7.75,0) node[sloped, pos=0.5, allow upside down]{\arrowIn};;
        \draw[black] (8.15,0)--(1.5+8,0) node[sloped, pos=0.5, allow upside down]{\arrowIn};;
        \draw[black] (7.93,0) circle (0.2cm);
        \draw[black] (-0.15+7.93,-0.15)--(0.15+7.93,0.15);
        \draw[black] (-0.15+7.93,0.15)--(0.15+7.93,-0.15);
        \filldraw (-1+7.8,0.2) circle (0pt) node[anchor=south]{$p$};
        \filldraw (1+8.2,0.2) circle (0pt) node[anchor=south]{$p$};
        \draw[black] (7.5,0.5) arc(225:-45:0.6)node[sloped, pos=0.5, allow upside down]{\arrowIn};;
        \draw[black] (7.5,0.5)--(7.93,0.2)--(8.35,0.5);
        \filldraw (8,1.5) circle (0pt) node[anchor=south]{$q$};
        \filldraw (10,0.5) circle (0pt) node[anchor=west]{$+$};
        \draw[black] (11.5,0)--(12.7,0) node[sloped, pos=0.5, allow upside down]{\arrowIn};;
        \draw[black] (13.3,0)--(14.5,0) node[sloped, pos=0.5, allow upside down]{\arrowIn};;
        \draw[black] (13,0) circle (0.3cm);
        \draw[black] (-0.21+13,-0.21)--(0.21+13,0.21);
        \draw[black] (-0.21+13,0.21)--(0.21+13,-0.21);
        \filldraw (-1+13,0.2) circle (0pt) node[anchor=south]{$p$};
        \filldraw (14,0.2) circle (0pt) node[anchor=south]{$p$};
        \end{tikzpicture}
    \caption{Feynman diagrams at the two-loop level for the two-point function.}
    \label{fig:2pt2loop}
\end{figure}
The diagrams in Figure.\ref{fig:2pt2loop} contribute the following,
\begin{equation}
    \begin{split}
        \circled{1} &: -\frac{R_{befg}R_{gaef}}{4} \int \frac{d^2q_1d^2q_2}{(2\pi)^4} \frac{i\slashed{p}}{p^2}\gamma^{\mu}\frac{i\slashed{q}_2}{q_2^2+i\epsilon}\gamma_{\nu}\frac{i\slashed{q}_1}{q_1^2+i\epsilon}\gamma_{\mu}\frac{i(\slashed{p}+\slashed{q}_1-\slashed{q}_2)}{(p+q_1-q_2)^2+i\epsilon}\gamma^{\nu}\frac{i\slashed{p}}{p^2} =0, \\
        &\text{since $\gamma^{\mu}(\cdots)\gamma_{\mu}=0$ where $(\cdots)$ consists of an odd number of gamma matrices.} \\
        \circled{2} &: \frac{R_{befg}R_{eagf}}{4}\int\frac{d^2q_1d^2q_2}{(2\pi)^4}\frac{i\slashed{p}}{p^2}\gamma^{\mu}\frac{i(\slashed{p}-\slashed{q}_1+\slashed{q}_2)}{(p-q_1+q_2)^2+i\epsilon}\gamma^{\nu}\frac{i\slashed{p}}{p^2}\Tr\bigg(\frac{i\slashed{q}_1}{q_1^2+i\epsilon}\gamma_{\mu}\frac{i\slashed{q}_2}{q_2^2+i\epsilon}\gamma_{\nu}\bigg)=0 , \\
        &\text{the numerator of the integrand $(p-q_1+q_2)^{\lambda}q_1^{\alpha}q_2^{\beta}$ is a sum of odd (after changes} \\
        &\text{of variables) functions of $q_1$ and/or $q_2$, thus integrating over $\{-\infty, \infty\}$ gives 0.} \\
        \circled{3} &: -\frac{i\delta^{(1)}_{beea}R_{beea}}{2}\int \frac{d^2q}{(2\pi)^2} \frac{i\slashed{p}}{p^2}\gamma^{\mu}\frac{i\slashed{q}}{q^2+i\epsilon}\gamma_{\mu}\frac{i\slashed{p}}{p^2} = 0 , \\
        & \text{since $\gamma^{\mu}(\cdots)\gamma_{\mu}=0$ where $\cdots$ are odd number of gamma matrices.} \\
        \circled{4} &: \frac{i\delta^{(1)}_{baee}R_{baee}}{2}\frac{i\slashed{p}}{p^2}\gamma^{\mu}\frac{i\slashed{p}}{p^2} \int \frac{d^2q}{(2\pi)^2}\Tr\bigg(\frac{i\slashed{q}}{q^2+i\epsilon}\gamma_{\mu}\bigg)  = 0, \\
        & \text{since the integrand is an odd function of $q$.} 
    \end{split}
\end{equation}
Indices $e,f,g$ are summed over $1, \dots, N$. Therefore, at the two-loop level the correction $\delta^{(2)}_a=0$.

\begin{figure}[H]
    \centering
    \begin{tikzpicture}[scale=0.5]
        \filldraw (-17,0) circle (0pt) node[anchor=west]{$G_2^{(4)}(x,y)$};
        \filldraw (-13.3,0) circle (0pt) node[anchor=west]{$=$};
        
        \draw[black] (1-10,-0.1) arc(-25:-155:0.55)node[sloped, pos=0.5, allow upside down]{\arrowIn};;
        \draw[black] (1-10,0.1) arc(25:155:0.55)node[sloped, pos=0.5, allow upside down]{\arrowIn};;
        \draw[black] (-10,-0.1) arc(-25:-155:0.55)node[sloped, pos=0.5, allow upside down]{\arrowIn};;
        \draw[black] (-10,0.1) arc(25:155:0.55)node[sloped, pos=0.5, allow upside down]{\arrowIn};;
        \draw[black] (-10,2)--(1-10,0.1) node[sloped, pos=0.5, allow upside down]{\arrowIn};;
        \draw[black] (-1-10,0.1)--(-10,2) node[sloped, pos=0.5, allow upside down]{\arrowIn};;
        \draw[black] (-10,-2)--(1-10,-0.1) node[sloped, pos=0.5, allow upside down]{\arrowIn};;
        \draw[black] (-1-10,-0.1)--(-10,-2) node[sloped, pos=0.5, allow upside down]{\arrowIn};;
        \filldraw (-0.5-10,-1.5) circle (0pt) node[anchor=east]{c};
        \filldraw (0.45-10,-1.43) circle (0pt) node[anchor=west]{d};
        \filldraw (-0.5-10,1.5) circle (0pt) node[anchor=east]{a};
        \filldraw (0.45-10,1.57) circle (0pt) node[anchor=west]{b};
        \filldraw (-8.3,0) circle (0pt) node[anchor=west]{$+$};
        
        \draw[black] (1-5.5,-0.1) arc(-25:-155:0.55)node[sloped, pos=0.5, allow upside down]{\arrowIn};;
        \draw[black] (-0.99-5.5,0.1) arc(155:25:0.55)node[sloped, pos=0.5, allow upside down]{\arrowIn};;
        \draw[black] (-5.5,-0.1) arc(-25:-155:0.55)node[sloped, pos=0.5, allow upside down]{\arrowIn};;
        \draw[black] (-5.5,0.1) arc(155:25:0.55)node[sloped, pos=0.5, allow upside down]{\arrowIn};;
        \draw[black] (1-5.5,0.1) arc(35:56:4);
        \draw[black] (-1-5.5,0.1) arc(145:127:4);
        \draw[black] (0-5.5,1.09) arc(239:120:0.56)node[sloped, pos=0.5, allow upside down]{\arrowIn};;
        \draw[black] (0-5.5,2.05) arc(65:-52:0.5)node[sloped, pos=0.5, allow upside down]{\arrowIn};;
        \draw[black] (0-5.5,-2)--(1-5.5,-0.1) node[sloped, pos=0.5, allow upside down]{\arrowIn};;
        \draw[black] (-1-5.5,-0.1)--(0-5.5,-2) node[sloped, pos=0.5, allow upside down]{\arrowIn};;
        \filldraw (-0.5-5.5,-1.5) circle (0pt) node[anchor=east]{c};
        \filldraw (0.45-5.5,-1.43) circle (0pt) node[anchor=west]{d};
        \filldraw (-0.5-5.5,1.5) circle (0pt) node[anchor=east]{a};
        \filldraw (0.45-5.5,1.57) circle (0pt) node[anchor=west]{b};
        \filldraw (0-3.7,0) circle (0pt) node[anchor=west]{$+$};
        
        \draw[black] (0-1,2)--(0.6-1,0.51) node[sloped, pos=0.5, allow upside down]{\arrowIn};;
        \draw[black] (-1-1,0.05)--(0-1,2) node[sloped, pos=0.5, allow upside down]{\arrowIn};;
        \draw[black] (0.55-1,-0.6)--(-1-1,0.05) node[sloped, pos=0.5, allow upside down]{\arrowIn};;
        \draw[black] (0-1,-2)--(0.65-1,-0.55) node[sloped, pos=0.5, allow upside down]{\arrowIn};;
        \draw[black] (-1-1,-0.05)--(0-1,-2) node[sloped, pos=0.5, allow upside down]{\arrowIn};;
        \draw[black] (0.53-1,0.47)--(-0.84-1,0.1) node[sloped, pos=0.5, allow upside down]{\arrowIn};;
        \draw[black] (0.6-1,0.51) arc(65:-52:0.6)node[sloped, pos=0.5, allow upside down]{\arrowIn};;
        \draw[black] (0.65-1,-0.55) arc(245:130:0.6)node[sloped, pos=0.5, allow upside down]{\arrowIn};;
        \filldraw (-0.5-1,-1.5) circle (0pt) node[anchor=east]{c};
        \filldraw (0.45-1,-1.43) circle (0pt) node[anchor=west]{d};
        \filldraw (-0.5-1,1.5) circle (0pt) node[anchor=east]{a};
        \filldraw (0.45-1,1.57) circle (0pt) node[anchor=west]{b};
        \filldraw (5.1-4.4,0) circle (0pt) node[anchor=west]{$+$};
        
        \draw[black] (0.6+3.5,0.51) arc(34:58:2.1);
        \draw[black] (-1+3.5,0.05) arc(148:129:4);
        \draw[black] (0+3.5,1.09) arc(239:120:0.56)node[sloped, pos=0.5, allow upside down]{\arrowIn};;
        \draw[black] (0+3.5,2.05) arc(65:-52:0.5)node[sloped, pos=0.5, allow upside down]{\arrowIn};;
        \draw[black] (-1+3.5,0.05)--(0.55+3.5,-0.6) node[sloped, pos=0.5, allow upside down]{\arrowIn};;
        \draw[black] (0+3.5,-2)--(0.65+3.5,-0.55) node[sloped, pos=0.5, allow upside down]{\arrowIn};;
        \draw[black] (-1+3.5,-0.05)--(0+3.5,-2) node[sloped, pos=0.5, allow upside down]{\arrowIn};;
        \draw[black] (0.53+3.5,0.47)--(-0.84+3.5,0.1) node[sloped, pos=0.5, allow upside down]{\arrowIn};;
        \draw[black] (0.71+3.5,-0.5) arc(-52:65:0.6)node[sloped, pos=0.5, allow upside down]{\arrowIn};;
        \draw[black] (0.65+3.5,-0.55) arc(245:130:0.6)node[sloped, pos=0.5, allow upside down]{\arrowIn};;
        \filldraw (-0.5+3.5,-1.5) circle (0pt) node[anchor=east]{c};
        \filldraw (0.45+3.5,-1.43) circle (0pt) node[anchor=west]{d};
        \filldraw (-0.5+3.5,1.5) circle (0pt) node[anchor=east]{a};
        \filldraw (0.45+3.5,1.57) circle (0pt) node[anchor=west]{b};
        \filldraw (5.2,0) circle (0pt) node[anchor=west]{$+$};
        
        \draw[black] (0+8,2)--(1+8,0.05) node[sloped, pos=0.5, allow upside down]{\arrowIn};;
        \draw[black] (-0.6+8,0.51)--(0+8,2) node[sloped, pos=0.5, allow upside down]{\arrowIn};;
        \draw[black] (0.84+8,0.1)--(-0.5+8,0.5) node[sloped, pos=0.5, allow upside down]{\arrowIn};;
        \draw[black] (0+8,-2)--(1+8,-0.05) node[sloped, pos=0.5, allow upside down]{\arrowIn};;
        \draw[black] (-0.6+8,-0.55)--(0+8,-2) node[sloped, pos=0.5, allow upside down]{\arrowIn};;
        \draw[black] (1+8,0.05)--(-0.46+8,-0.57) node[sloped, pos=0.5, allow upside down]{\arrowIn};;
        \draw[black] (-0.5+8,0.5) arc(56:-65:0.6)node[sloped, pos=0.5, allow upside down]{\arrowIn};;
        \draw[black] (-0.67+8,-0.5) arc(235:119:0.6)node[sloped, pos=0.5, allow upside down]{\arrowIn};;
        \filldraw (-0.5+8,-1.5) circle (0pt) node[anchor=east]{c};
        \filldraw (0.45+8,-1.43) circle (0pt) node[anchor=west]{d};
        \filldraw (-0.5+8,1.5) circle (0pt) node[anchor=east]{a};
        \filldraw (0.45+8,1.57) circle (0pt) node[anchor=west]{b};
        \filldraw (5+4.7,0) circle (0pt) node[anchor=west]{$+$};

        \draw[black] (1+12.5,0.05) arc(33:55:3.9);
        \draw[black] (-0.63+12.5,0.5) arc(144:134:4);
        \draw[black] (0+12.5,1.09) arc(239:120:0.56)node[sloped, pos=0.5, allow upside down]{\arrowIn};;
        \draw[black] (0+12.5,2.05) arc(65:-52:0.5)node[sloped, pos=0.5, allow upside down]{\arrowIn};;
        \draw[black] (0.84+12.5,0.1)--(-0.5+12.5,0.47) node[sloped, pos=0.5, allow upside down]{\arrowIn};;
        \draw[black] (0+12.5,-2)--(1+12.5,-0.05) node[sloped, pos=0.5, allow upside down]{\arrowIn};;
        \draw[black] (-0.6+12.5,-0.55)--(0+12.5,-2) node[sloped, pos=0.5, allow upside down]{\arrowIn};;
        \draw[black] (-0.46+12.5,-0.6)--(1+12.5,0.05) node[sloped, pos=0.5, allow upside down]{\arrowIn};;
        \draw[black] (-0.5+12.5,0.47) arc(56:-65:0.6)node[sloped, pos=0.5, allow upside down]{\arrowIn};;
        \draw[black] (-0.63+12.5,0.5) arc(119:235:0.6)node[sloped, pos=0.5, allow upside down]{\arrowIn};;
        \filldraw (-0.5+12.5,-1.5) circle (0pt) node[anchor=east]{c};
        \filldraw (0.45+12.5,-1.43) circle (0pt) node[anchor=west]{d};
        \filldraw (-0.5+12.5,1.5) circle (0pt) node[anchor=east]{a};
        \filldraw (0.45+12.5,1.57) circle (0pt) node[anchor=west]{b};
        \filldraw (-10.5,-5) circle (0pt) node[anchor=west]{$+$};

        \draw[black] (2-8.7,-0.2-5) arc(-35:-145:1.23)node[sloped, pos=0.5, allow upside down]{\arrowIn};;
        \draw[black] (2-8.7,0.2-5) arc(35:145:1.23)node[sloped, pos=0.5, allow upside down]{\arrowIn};;
        \draw[black] (1-8.7,2.1-5)--(2-8.7,0.2-5) node[sloped, pos=0.5, allow upside down]{\arrowIn};;
        \draw[black] (0-8.7,0.2-5)--(1-8.7,2.1-5) node[sloped, pos=0.5, allow upside down]{\arrowIn};;
        \draw[black] (1-8.7,-2.1-5)--(2-8.7,-0.2-5) node[sloped, pos=0.5, allow upside down]{\arrowIn};;
        \draw[black] (0-8.7,-0.2-5)--(1-8.7,-2.1-5) node[sloped, pos=0.5, allow upside down]{\arrowIn};;
        \draw[black] (-8.7,0-5) circle (0.2cm);
        \draw[black] (-0.15-8.7,-0.15-5)--(0.15-8.7,0.15-5);
        \draw[black] (-0.15-8.7,0.15-5)--(0.15-8.7,-0.15-5);
        \filldraw (0.5-8.7,-1.5-5) circle (0pt) node[anchor=east]{c};
        \filldraw (1.45-8.7,-1.43-5) circle (0pt) node[anchor=west]{d};
        \filldraw (0.5-8.7,1.5-5) circle (0pt) node[anchor=east]{a};
        \filldraw (1.45-8.7,1.57-5) circle (0pt) node[anchor=west]{b};
        \filldraw (2-8,0-5) circle (0pt) node[anchor=west]{$+$};

        \draw[black] (6-8.2,-0.2-5) arc(-35:-145:1.23)node[sloped, pos=0.5, allow upside down]{\arrowIn};;
        \draw[black] (4-8.2,0.2-5) arc(145:35:1.22)node[sloped, pos=0.5, allow upside down]{\arrowIn};;
        \draw[black] (6-8.2,0.2-5) arc(35:56:4);
        \draw[black] (4-8.2,0.2-5) arc(145:127:4);
        \draw[black] (5-8.2,1.19-5) arc(239:120:0.56)node[sloped, pos=0.5, allow upside down]{\arrowIn};;
        \draw[black] (5-8.2,2.15-5) arc(65:-52:0.5)node[sloped, pos=0.5, allow upside down]{\arrowIn};;
        \draw[black] (5-8.2,-2.1-5)--(6-8.2,-0.2-5) node[sloped, pos=0.5, allow upside down]{\arrowIn};;
        \draw[black] (4-8.2,-0.2-5)--(5-8.2,-2.1-5) node[sloped, pos=0.5, allow upside down]{\arrowIn};;
        \draw[black] (-4.2,0-5) circle (0.2cm);
        \draw[black] (-0.15-4.2,-0.15-5)--(0.15-4.2,0.15-5);
        \draw[black] (-0.15-4.2,0.15-5)--(0.15-4.2,-0.15-5);
        \filldraw (4.5-8.2,-1.5-5) circle (0pt) node[anchor=east]{c};
        \filldraw (5.45-8.2,-1.43-5) circle (0pt) node[anchor=west]{d};
        \filldraw (4.5-8.2,1.5-5) circle (0pt) node[anchor=east]{a};
        \filldraw (5.45-8.2,1.57-5) circle (0pt) node[anchor=west]{b};
        \filldraw (-1.5,0-5) circle (0pt) node[anchor=west]{$+$};

        \draw[black] (2+0.3,-0.2-5) arc(-35:-145:1.23)node[sloped, pos=0.5, allow upside down]{\arrowIn};;
        \draw[black] (2+0.3,0.2-5) arc(35:145:1.23)node[sloped, pos=0.5, allow upside down]{\arrowIn};;
        \draw[black] (1+0.3,2.1-5)--(2+0.3,0.2-5) node[sloped, pos=0.5, allow upside down]{\arrowIn};;
        \draw[black] (0+0.3,0.2-5)--(1+0.3,2.1-5) node[sloped, pos=0.5, allow upside down]{\arrowIn};;
        \draw[black] (1+0.3,-2.1-5)--(2+0.3,-0.2-5) node[sloped, pos=0.5, allow upside down]{\arrowIn};;
        \draw[black] (0+0.3,-0.2-5)--(1+0.3,-2.1-5) node[sloped, pos=0.5, allow upside down]{\arrowIn};;
        \draw[black] (2.3,0-5) circle (0.2cm);
        \draw[black] (-0.15+2.3,-0.15-5)--(0.15+2.3,0.15-5);
        \draw[black] (-0.15+2.3,0.15-5)--(0.15+2.3,-0.15-5);
        \filldraw (0.5+0.3,-1.5-5) circle (0pt) node[anchor=east]{c};
        \filldraw (1.45+0.3,-1.43-5) circle (0pt) node[anchor=west]{d};
        \filldraw (0.5+0.3,1.5-5) circle (0pt) node[anchor=east]{a};
        \filldraw (1.45+0.3,1.57-5) circle (0pt) node[anchor=west]{b};
        \filldraw (3,0-5) circle (0pt) node[anchor=west]{$+$};

        \draw[black] (6+0.7,-0.2-5) arc(-35:-145:1.23)node[sloped, pos=0.5, allow upside down]{\arrowIn};;
        \draw[black] (4+0.7,0.2-5) arc(145:35:1.22)node[sloped, pos=0.5, allow upside down]{\arrowIn};;
        \draw[black] (6+0.7,0.2-5) arc(35:56:4);
        \draw[black] (4+0.7,0.2-5) arc(145:127:4);
        \draw[black] (5+0.7,1.19-5) arc(239:120:0.56)node[sloped, pos=0.5, allow upside down]{\arrowIn};;
        \draw[black] (5+0.7,2.15-5) arc(65:-52:0.5)node[sloped, pos=0.5, allow upside down]{\arrowIn};;
        \draw[black] (5+0.7,-2.1-5)--(6+0.7,-0.2-5) node[sloped, pos=0.5, allow upside down]{\arrowIn};;
        \draw[black] (4+0.7,-0.2-5)--(5+0.7,-2.1-5) node[sloped, pos=0.5, allow upside down]{\arrowIn};;
        \draw[black] (6.7,0-5) circle (0.2cm);
        \draw[black] (-0.15+6.7,-0.15-5)--(0.15+6.7,0.15-5);
        \draw[black] (-0.15+6.7,0.15-5)--(0.15+6.7,-0.15-5);
        \filldraw (4.5+0.7,-1.5-5) circle (0pt) node[anchor=east]{c};
        \filldraw (5.45+0.7,-1.43-5) circle (0pt) node[anchor=west]{d};
        \filldraw (4.5+0.7,1.5-5) circle (0pt) node[anchor=east]{a};
        \filldraw (5.45+0.7,1.57-5) circle (0pt) node[anchor=west]{b};
        \filldraw (-1.5+8.9,0-5) circle (0pt) node[anchor=west]{$+$};
        
        \draw[black] (4.8+5.2,0.2-5) arc(225:120:1.05)node[sloped, pos=0.5, allow upside down]{\arrowIn};;
        \draw[black] (5+5.2,1.85-5) arc(65:-51:1)node[sloped, pos=0.5, allow upside down]{\arrowIn};;
        \draw[black] (10.2,0-5) circle (0.3cm);
        \draw[black] (-0.21+10.2,-0.21-5)--(0.21+10.2,0.21-5);
        \draw[black] (-0.21+10.2,0.21-5)--(0.21+10.2,-0.21-5);
        \draw[black] (13.05-2.8,-1.85-5) arc(-60:48:1.03)node[sloped, pos=0.5, allow upside down]{\arrowIn};;
        \draw[black] (13-0.2-2.8,-0.2-5) arc(130:245:1)node[sloped, pos=0.5, allow upside down]{\arrowIn};;
        \filldraw (4.5+5.2,-1.5-5) circle (0pt) node[anchor=east]{c};
        \filldraw (5.45+5.2,-1.43-5) circle (0pt) node[anchor=west]{d};
        \filldraw (4.5+5.2,1.5-5) circle (0pt) node[anchor=east]{a};
        \filldraw (5.45+5.2,1.57-5) circle (0pt) node[anchor=west]{b};
        \end{tikzpicture}
    \caption{Feynman diagrams at the two-loop level for the four-point function.}
    \label{fig:4pt2loop}
\end{figure}

At the two-loop level, 21600 different Wick contractions are reduced to 96 distinct Feynman diagrams. This is still a huge number to read off correlation functions from. Applying tricks presented in appendix.\ref{sec:appendixB}, we can drop 90 vanishing diagrams and keep only 6 of them. Using results from the current-current correlation function at the one-loop level and bringing in Eq.(\ref{eq:4pt1loop}), we easily obtain contributions from the diagrams in the second row of Figure.\ref{fig:4pt2loop}, 
\begin{equation}
    \begin{split}
        \circled{1}_2 &= \frac{iP\Gamma   }{4\pi}\sum_{e, f=1}^N \delta^{(1)}_{eafc}R_{eafc}R_{bedf}= \frac{iP\Gamma   }{8\pi^2\epsilon^2}\sum_{e,f,g,h=1}^N(R_{egfh}R_{gahc} - R_{eghc}R_{gafh})R_{bedf}, \\
        \circled{2}_2 &= -\frac{iP\Gamma   }{4\pi}\sum_{e, f=1}^N \delta^{(1)}_{befc}R_{eadf}R_{befc}= -\frac{iP\Gamma   }{8\pi^2\epsilon^2}\sum_{e,f,g,h=1}^N (R_{bgfh}R_{gehc} - R_{bghc}R_{gefh})R_{eadf} ,\\
        \circled{3}_2 &= \frac{iP\Gamma   }{4\pi}\sum_{e, f=1}^N \delta^{(1)}_{bedf}R_{eafc}R_{bedf}=\frac{iP\Gamma   }{8\pi^2\epsilon^2}\sum_{e,f,g,h=1}^N(R_{bgdh}R_{gehf} - R_{bghf}R_{gedh})R_{eafc} ,\\
        \circled{4}_2 &= -\frac{iP\Gamma   }{4\pi}\sum_{e, f=1}^N \delta^{(1)}_{eadf}R_{eadf}R_{befc}= -\frac{iP\Gamma   }{8\pi^2\epsilon^2}\sum_{e,f,g,h=1}^N(R_{egdh}R_{gahf} - R_{eghf}R_{gadh})R_{befc},\\
        \circled{5}_2 &= -\frac{iP\Gamma}{2}\tilde{\delta}^{(2)}_{badc}R_{badc}.
    \end{split}
\end{equation}
Diagrams in the first row of Figure.\ref{fig:4pt2loop} contribute
\begin{equation}
    \begin{split}
        \circled{1}_1 &= -\frac{iP\Gamma  }{8\pi^2\epsilon^2}\sum_{e,f,g,h=1}^N R_{eafc}R_{bgdh}R_{gehf} ,\hspace{10mm}
        \circled{2}_1 = -\frac{iP\Gamma  }{8\pi^2\epsilon^2}\sum_{e,f,g,h=1}^N R_{eadf}R_{bghc}R_{gefh}, \\
        \circled{3}_1 &= \frac{iP\Gamma  }{8\pi^2\epsilon^2}\sum_{e,f,g,h=1}^N \frac{1}{2}R_{eafc}R_{bghf}R_{gedh}, \hspace{10mm}
        \circled{4}_1 = \frac{iP\Gamma  }{8\pi^2\epsilon^2}\sum_{e,f,g,h=1}^N \frac{1}{2}R_{eahf}R_{bgfc}R_{gedh} , \\
        \circled{5}_1 &= \frac{iP\Gamma  }{8\pi^2\epsilon^2}\sum_{e,f,g,h=1}^N \frac{1}{2}R_{eafg}R_{bhdf}R_{hegc},  \hspace{10.5mm}
        \circled{6}_1 = \frac{iP\Gamma  }{8\pi^2\epsilon^2}\sum_{e,f,g,h=1}^N \frac{1}{2}R_{eadf}R_{bgfh}R_{gehc} .
    \end{split}
\end{equation}

We see that each integral over $q_i$ gives a divergence of $1/\epsilon$, so each diagram contributes a second-order infinity to the correction. Up to two-loop level, $\tilde{Z}_{abcd} = (Z_aZ_bZ_cZ_d)^{1/2}\big(1+\delta^{(1)}_{abcd}+\delta^{(2)}_{abcd}\big) = 1+\tilde{\delta}^{(1)}_{abcd}+\tilde{\delta}^{(2)}_{abcd}$, we have shown that $\delta^{(1)}_a=\delta^{(2)}_a=0$ and $\tilde{\delta}^{(1)}_{abcd}=\delta^{(1)}_{abcd}$, it is trivial that $\tilde{\delta}^{(2)}_{abcd}=\delta^{(2)}_{abcd}$. The correction to the beta function at the 2-loop level is given by
\begin{equation}
    \begin{split}
        \delta^{(2)}_{abcd}R_{abcd}&= \frac{1}{4\pi^2\epsilon^2}\sum_{e,f,g,h=1}^N \bigg(R_{aecf}R_{gbhd}R_{egfh} + R_{aefd}R_{gbch}R_{eghf} - \frac{1}{2} R_{aecf}R_{gbfh}R_{eghd}  \\
        &\hspace{10mm} -\frac{1}{2} R_{aefg}R_{hbcf}R_{ehgd} - \frac{1}{2}R_{aefd}R_{gbhf}R_{egch} -\frac{1}{2} R_{aefg}R_{hbgd}R_{ehcf}\bigg) \equiv \frac{\mathcal{R}^{(2)}_{abcd}}{4\pi^2\epsilon^2}.
    \end{split}
\end{equation}

\section{The $\beta$-functions and Constraints on Coupling Constants}\label{sec:betafunction}
\hspace{6mm}Recall the definition of the $\beta$-function, 
\begin{equation}
    \beta_{abcd}(R_{abcd}(\mu)) \equiv \mu \frac{\partial R_{abcd}}{\partial \mu} ,
\end{equation}
where $R_{abcd}(\mu)$ is the renormalized coupling constant which depends on the renormalization scale $\mu$. $\mu$, with mass dimension $[\mu]=1$, is a parameter introduced to carry out dimensional regularization, which is not a physical parameter but an artifice of the renormalization procedure. To regularize the divergence in the loop diagrams, we perform dimensional regularization and take $d=2-\epsilon$ with limit $\epsilon \rightarrow 0$. Now we count the mass dimension of each bare quantity. Since the action $\mathcal{S}=\int d^d x \mathcal{L}$ is dimensionless, we have $[\mathcal{L}]=d$, then
\begin{equation}
    \begin{split}
        [\psi^a_0] = \frac{d-1}{2} = \frac{1-\epsilon}{2}, \hspace{3mm} \text{and} \hspace{3mm} [R^0_{abcd}]=d-4[\psi_0^a] = \epsilon .
    \end{split}
\end{equation}
The bare couplings ought to be independent of $\mu$, therefore we need to replace $R^0_{abcd}$ with $\mu^{\epsilon}R^0_{abcd}$ in the Lagrangian to make it a dimensionless the following equation
\begin{equation}
    \begin{split}
        \frac{\partial R^0_{abcd}}{\partial \mu} = \frac{\partial}{\partial \mu} (\mu^{\epsilon}Z_{abcd}R_{abcd}) = 0.
    \end{split}
\end{equation}
Expand the right hand side and multiply $\mu^{1-\epsilon}Z_{abcd}^{-1}$ on both sides, we get the expression of the $\beta$-function for $R_{abcd}$,
\begin{equation}\label{eq:betaabcd}
    \begin{split}
        \mu\frac{\partial R_{abcd}}{\partial \mu} = -\epsilon R_{abcd} - \frac{R_{abcd}}{Z_{abcd}}\bigg(\mu\frac{\partial Z_{abcd}}{\partial \mu}\bigg) .
    \end{split}
\end{equation}

\begin{equation}\label{eq:Zabcd}
    \begin{split}
        \mu\frac{\partial Z_{abcd}}{\partial \mu} = \mu \frac{\partial \delta_{abcd}}{\partial \mu} = -\frac{\mathcal{R}_{abcd}}{2\pi\epsilon R^2_{abcd}}\bigg(\mu\frac{\partial R_{abcd}}{\partial \mu}\bigg) + \frac{1}{2\pi\epsilon R_{abcd}}\bigg(\mu\frac{\partial \mathcal{R}_{abcd}}{\partial \mu}\bigg).
    \end{split}
\end{equation}
Bring Eq.(\ref{eq:Zabcd}) into Eq.(\ref{eq:betaabcd}) we get

\begin{equation}
    \begin{split}
        \beta_{abcd} &= -\epsilon R_{abcd} - \frac{\mathcal{R}_{abcd}}{2\pi} - \frac{1}{2\pi\epsilon} \bigg(\mu\frac{\partial \mathcal{R}_{abcd}}{\partial \mu}\bigg) 
        + \mathcal{O}\big(R^4_{abcd}\big) \\
        &\equiv \beta^{(0)}_{abcd} + \beta^{(1)}_{abcd} + \beta^{(2)}_{abcd} + \mathcal{O}\big(R^4_{abcd}\big),
    \end{split}
\end{equation}
where $\mathcal{R}_{abcd} \equiv \mathcal{R}^{(1)}_{abcd} + \frac{1}{2\pi}\mathcal{R}^{(2)}_{abcd}$ is a function of coupling constants $R_{abcd}$'s to the second order in the first term and third order in the second term. $\mathcal{R}^{(k)}_{abcd}$ is defined as $\delta^{(k)}_{abcd}R_{abcd}\sim \mathcal{R}^{(k)}_{abcd}$ up to a coefficient that depends on $\epsilon$ and is determined by cancelling k-loop level divergence in the current-current correlation function. The beta functions are just combinations of the coefficients of the divergent logarithms and they are given by
\begin{equation}
    \begin{split}
        \beta^{(0)}_{abcd} & = - \epsilon R_{abcd}, \hspace{20mm}
        \beta^{(1)}_{abcd}  = \frac{\mathcal{R}^{(1)}_{abcd}}{2\pi}, \hspace{20mm}
        \beta^{(2)}_{abcd}  = \frac{2\mathcal{R}^{(2)}_{abcd}}{(2\pi)^2}.
    \end{split}
\end{equation}

As we can see in the calculations in section \ref{subsec:2loop}, the two-loop beta function is obtained by replacing one of the vertices with the one-loop counterterms, therefore vanishing beta functions at the one-loop level automatically yield vanishing beta function at the two-loop level. Setting the one-loop corrections to the $\beta$-function to zero, we get the following constraints that
\begin{equation}\label{eq:betafcncondition}
    \begin{split}
        \mathcal{R}^{(1)}_{abcd} &\equiv \sum_{e, f=1}^N(R_{aecf}R_{ebfd} - R_{aefd}R_{ebcf}) = 0.
    \end{split}
\end{equation}
To find the most general constraints on the coupling constants $R_{abcd}$'s such that the model is a CFT, we look for solutions to Eq.(\ref{eq:betafcncondition}). An obvious solution takes the form of
\begin{equation}\label{eq:soln}
    R_{abcd}^2=R_{abab}R_{cdcd},
\end{equation}
this solution is sufficient but not necessary, and there are solutions beyond this form. In either the Majorana or the Dirac cases, the degrees of freedom (DOFs) $D$ in the couplings grow as $\mathcal{O}(N^4)$. The beta function constraints remove some DOFs, to find the remaining DOFs one needs to analytically solve the system of Eq.(\ref{eq:betafcncondition}).

In the Dirac cases, the DOFs are larger due to fewer symmetry properties in the indices, which is given by $D = \frac{1}{2}N^2(N^2+1)$. One can verify that Eq.(\ref{eq:soln}) is indeed a root of the beta functions, but there are other roots to Eq.(\ref{eq:betafcncondition}) beyond this form for any $N\geq 2$, this is numerically verified. The form in Eq.(\ref{eq:soln}) requires the number of DOFs in couplings to be reduced to $N^2$, which is not the case even for small $N$. For example, consider the case of $N=2$ Dirac fermions, there are $10$ independent couplings $R_{abcd}$'s and the vanishing beta function conditions has an analytic solution with minimal zero couplings,
\begin{equation}\label{eq:2diracsoln}
    \begin{split}
        R_{1122} &=\frac{1}{2R_{1212}}\Big[R_{1111}R_{1212}+R_{1212}R_{2222}+2R_{1112}R_{1222}-R_{1112}^2-R_{1222}^2\Big], \\
        R_{1112} &= \frac{1}{R_{1212}}\Big[R_{1222}R_{2121}+R_{1221}(R_{1121}-R_{2122})\Big], \\
        R_{1221}^2 &= R_{1212}R_{2121}, 
    \end{split}
\end{equation}
such a solution removes only $3$ DOFs with $7$ DOFs remaining, and $7>2^2$. For $N\geq 3$, analytic solutions are difficult to find, but we can solve the system numerically. Numerical examples of a general solution to vanishing beta functions for $N=2,3$, and $4$ are shown in Tables \ref{table:2dirac}, \ref{table:3dirac}, and \ref{table:4dirac1},\ref{table:4dirac2} in appendix \ref{sec:appendixC}, respectively. 

\begin{figure}
    \begin{minipage}{0.45\textwidth}
    \centering
    \includegraphics[scale = 0.67]{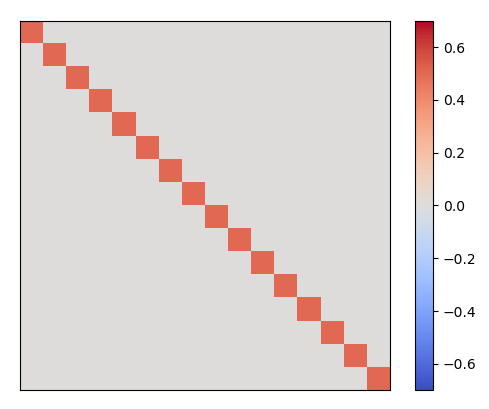}
    \caption{Color map of couplings in the Abelian Thirring model with $N=4$ Dirac fermions, where $R_{abcd} = g\delta_{ab}\delta_{cd}$ with $g=1/2$.}
    \label{fig:abelian}
    \end{minipage}
    \hspace{5mm}
    \begin{minipage}{0.45\textwidth}
    \centering
    \includegraphics[scale = 0.67]{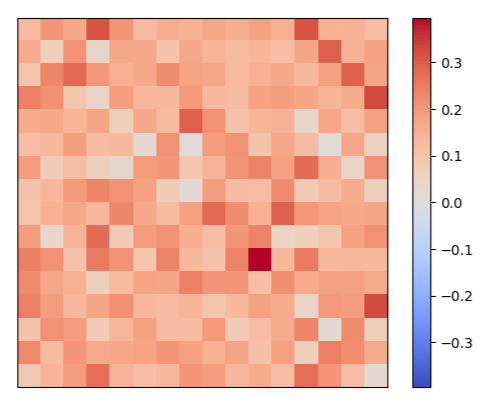}
    \caption{An example solution to the beta functions up to the two-loop level for N=4 Dirac fermions that has no obvious pattern.}
    \label{fig:4diracrandom}
\end{minipage}
\end{figure}

The number of couplings grows as $\mathcal{O}(N^4)$, which quickly becomes a large amount of data to analyze. To visualize the solutions, we create color maps using the numerical results and we brief the coloring procedure in the following. The coupling constants $R_{abcd}$'s can be mapped into a $N\times N$ matrix $\mathbf{R}$ of $N\times N$ matrices $\mathbf{R}_{ab}$,
\begin{equation}\label{eq:matrix}
    \begin{split}
        \mathbf{R} = \big(\mathbf{R}_{ab}\big) \hspace{5mm} \text{where} \hspace{5mm} \mathbf{R}_{ab} = \big(R_{ab;cd}\big) \hspace{5mm} \text{with} \hspace{5mm} 1\leq a,b,c,d\leq N,
    \end{split}
\end{equation}
and we define $R_{ab;cd}\equiv R_{abcd}$. More explicitly, we map the couplings in the following way
\begin{equation}
    \mathbf{R} = \begin{pmatrix}
        \begin{pmatrix}
            R_{1111} & \cdots & R_{111N} \\
            \vdots & \ddots & \vdots \\
            R_{11N1} & \cdots & R_{11NN} 
        \end{pmatrix} & \cdots &
        \begin{pmatrix}
            R_{1N11} & \cdots & R_{1N1N} \\
            \vdots & \ddots & \vdots \\
            R_{1NN1} & \cdots & R_{1NNN}
        \end{pmatrix} \\
        \vdots & \ddots & \vdots \\
        \begin{pmatrix}
            R_{N111} & \cdots & R_{N11N} \\
            \vdots & \ddots & \vdots \\
            R_{N1N1} & \cdots & R_{N1NN} 
        \end{pmatrix} & \cdots &
        \begin{pmatrix}
            R_{NN11} & \cdots & R_{NN1N} \\
            \vdots & \ddots & \vdots \\
            R_{NNN1} & \cdots & R_{NNNN}
        \end{pmatrix}
    \end{pmatrix}.
\end{equation}
Then we impose the symmetry properties on the couplings to remove some DOFs. Replacing the couplings by their corresponding numerical results, we get a $N^2 \times N^2$ table of data. Then we assign a color scale to values with gray for numerical zero, red for positive, and blue for negative, the darker the color, the larger the absolute value. Now we fill in the table with colors corresponding to numerical results of $R_{abcd}$'s to obtain a color map of a set of solutions to the couplings. Consider the Abelian Thirring model as a simple example, where $R_{abcd}=g\delta_{ab}\delta_{cd}$, its color map is shown in Figure.\ref{fig:abelian}.

\begin{figure}
\begin{minipage}{0.48\textwidth}
    \centering
    \includegraphics[scale = 0.6]{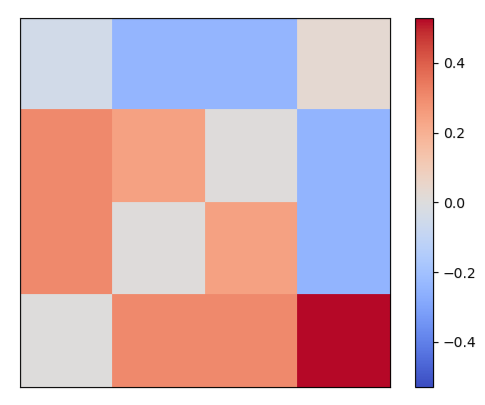}
    \caption{A solution for N=2 Dirac fermions.}
    \label{fig:2dirac}
\end{minipage}
\begin{minipage}{0.48\textwidth}
    \centering
    \includegraphics[scale = 0.6]{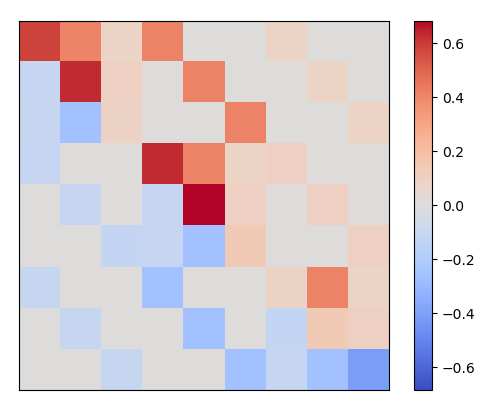}
    \caption{A solution for N=3 Dirac fermions.}
    \label{fig:3dirac}
\end{minipage}
\end{figure}

\begin{figure}
\begin{minipage}{0.48\textwidth}
    \centering
    \includegraphics[scale = 0.6]{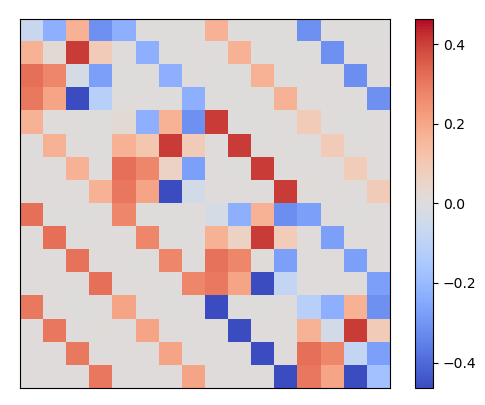}
    \caption{A solution for N=4 Dirac fermions.}
    \label{fig:4dirac}
\end{minipage}
\begin{minipage}{0.48\textwidth}
    \centering
    \includegraphics[scale = 0.6]{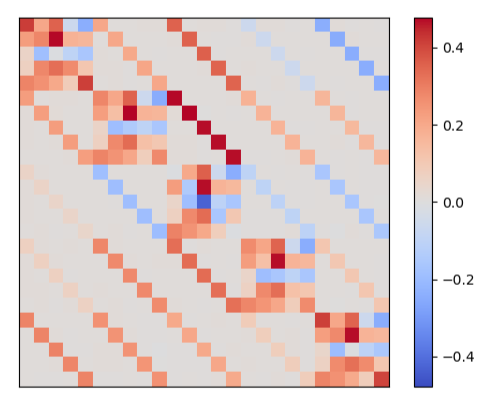}
    \caption{A solution for N=5 Dirac fermions.}
    \label{fig:5dirac}
\end{minipage}
\end{figure}

\begin{figure}
\begin{minipage}{0.48\textwidth}
    \centering
    \includegraphics[scale = 0.6]{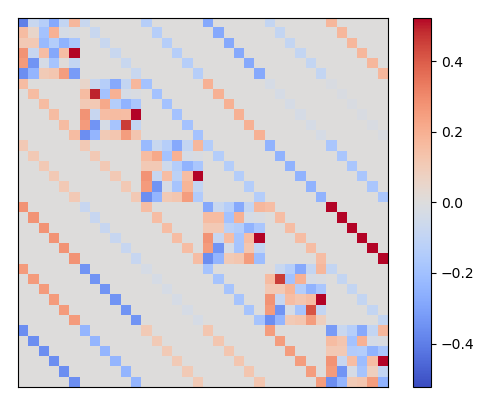}
    \caption{A solution for N=6 Dirac fermions.}
    \label{fig:6dirac}
\end{minipage}
\begin{minipage}{0.48\textwidth}
    \centering
    \includegraphics[scale = 0.6]{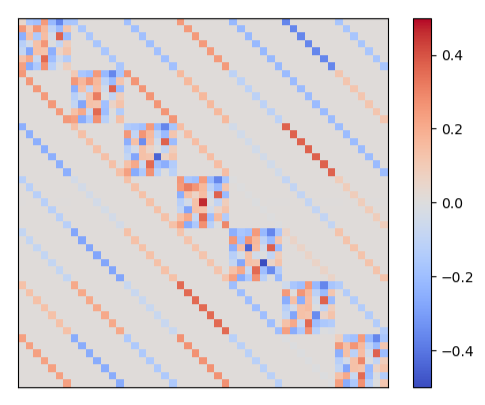}
    \caption{A solution for N=7 Dirac fermions.}
    \label{fig:7dirac}
\end{minipage}
\end{figure}

Following this procedure, we generate color maps for $N=2$ to $N=7$ Dirac fermions, shown in Figures.\ref{fig:2dirac}-\ref{fig:7dirac}. We see a pattern of solutions in these color maps as $N$ gets larger. Diagonal submatrices $\mathbf{R}_{aa}$'s share a similar pattern for $1\leq a\leq N$. Diagonal terms in off-diagonal submatrices are a few orders stronger than the rest couplings, creating diagonal strips in the color maps. These locations in the maps correspond to couplings in $R_{aabb}$ form with $1\leq a,b\leq N$, and the ones in $R_{abcc}$ form with $1\leq a,b,c\leq N$ and $a\neq b$. These couplings are more intense (i.e. have larger numerical values) than the rest, as the rest couplings are mapped to gray. Such a solution is more general than one in Eq.(\ref{eq:soln}) form. In addition, couplings in $R_{abcc}$ form with the same pair of $a,b$, for example, $R_{1233}$ and $R_{1244}$, have similar magnitude, as we see in the maps that diagonal terms in the same off-diagonal submatrices share very similar color, but diagonal terms in different off-diagonal submatrices have different colors. If solutions of the couplings completely concentrate on these patterns, the diagonal terms $R_{aabb}$'s give $\frac{1}{2}N(N+1)$ DOFs, and the $N$ diagonal terms $R_{abcc}$'s in any off-diagonal submatrices gives DOFs to $N^2(N-1)$, then in total the DOFs for a theory of $N$ Dirac fermions is given by
\begin{equation}
    D_N = N^3 - \frac{1}{2}N^2+\frac{1}{2}N.
\end{equation}
Bringing $N=1$ and $N=2$ into this formula, we get $D_1=1$ for the original Thrring model, and $D_2=7$ which matches with the DOFs given by the symbolic solution in Eq.(\ref{eq:2diracsoln}). 

We note that for $N=4$ Dirac fermions some numerical solutions are more random than the one shown in Figure.\ref{fig:4dirac} and do not have an obvious pattern as for other values of $N$, an example is shown in Figure.\ref{fig:4diracrandom}. A theory with $N=4$ Majorana fermions also behaves oddly where multiple patterns appear in the color map. We include these color maps and a discussion in appendix \ref{sec:appendixD}.

In Majorana cases, $D= \frac{1}{8}N^2(N-1)^2+\frac{1}{4}N(N-1)$. When $N=1$ it is a free theory as no interaction is allowed by the Grassmannian nature of fermions. When $N=2$ the theory has only one nonzero coupling $R_{1212}$ and the beta function vanishes automatically. $N=3$ theory has $6$ DOFs by symmetry properties in the indices, and the beta functions have an analytic solution with minimal zero couplings, which is given by
\begin{equation}
    R_{1213}^2 = R_{1212}R_{1313}, \hspace{10mm} R_{1223}^2=R_{1212}R_{2323}, \hspace{10mm} R_{1323}^2=R_{1313}R_{2323},
\end{equation}
where $3$ DOFs are removed and only $3$ remain. One can also numerically check that for $N=1,2,3$, there are no other solutions to the one-loop beta function other than the one given in Eq.(\ref{eq:soln}). However, for $N\geq 4$ analytically solving the system becomes difficult. Numerical results show that there are indeed additional solutions that are not in the above form. The DOFs of the most general solution grow at least as $\mathcal{O}(N^2)$. This bound is given by the solution form in Eq.(\ref{eq:soln}), it reduces the DOFs to $N$ choose $2$ in the Majorana case, i.e. $C_N^2 = \frac{1}{2}N(N-1)$ for $\forall N\geq 2$. See appendix \ref{sec:appendixD} for color map samples of numerical solutions. However, for Majorana fermions color maps are not as helpful in DOF reduction as in the Dirac cases. As one can see in the figures in the appendix, all couplings that survive after symmetry properties are imposed are turned on, and no obvious relation between the couplings manifests. A theory with $N=4$ Majorana fermions is odd compared to other numbers of fermions because multiple patterns appear in the color maps, see appendix \ref{sec:appendixD} for more details.

Even though exact symbolic solutions have not yet been determined, the calculation successfully reproduces some of the known interacting fermionic CFTs. The original Thirring model of $N=1$ Dirac fermion is a CFT that is equivalent to a free bosonic CFT compactified on a circle \cite{Abdalla:1991vua}. The Thirring coupling $g$ relates to the compact free boson coupling which is the radius $R$ of the circle in the following way
\begin{equation}\label{eq:b&f}
    \frac{R^2}{4} = \frac{\pi}{\pi+g} \hspace{1mm},
\end{equation}
When the Thirring coupling vanishes, i.e. $g=0$, the Thirring model becomes simply the free Dirac fermion, and the radius of the equivalent compact free boson corresponding to $g=0$ is $R=2$, which, through the T-duality, is dual to the $R=1$ case (recall that we have chosen $\alpha'=2$). Therefore, the above relation reduces the duality we discussed in the previous section. The $N=1$ Majorana fermion theory is a free fermionic theory since the interaction $(\overline{\psi}\gamma^{\mu}\psi)^2$ does not exist in this case due to the Grassmannian nature of fermions, and a free fermionic QFT is conformally invariant. When the theory has $N=2$ Majorana fermions, one can form a complex fermion. In general, this requires the Majorana fermions to have the same mass, in the massless case the requirement is satisfied. This model is a CFT, equivalent to a free boson compactified on a $\mathbb{Z}_2$ orbifold through bosonization \cite{DiFrancesco:1997nk}. In addition, letting $R_{abcd} = g\delta_{ab}\delta_{cd}$ in the generalized Thirring model yields Abelian Thirring model, which is also a CFT of an arbitrary number of $N$ Dirac fermions with an interaction of the abelian currents. The above theories have vanishing $\beta$-function at the one and two-loop level because they are CFTs, and this is verified using Mathematica.

\section{Conclusions}

\hspace{6mm}Motivated by the quantum gravity considerations outlined in \cite{hilbertbundles} we have revisited the question of lines of fixed points emanating from models of free fermions in $1 + 1$ dimensions.  Through two loop order we have found a new family of possibilities, in addition to the well known Abelian Thirring models.  In addition there are some numerical indications of additional solutions.  Of course, vanishing of the beta functions through two loop order means that the three loop contributions are scheme independent and might well be non-zero, so our work is only the first step in establishing the existence of true fixed lines. We reserve the difficult task of higher loop calculations to future work.  

 We note however that for potential applications to quantum gravity we are primarily interested in models with a very large number of fermion fields. In this case it is possible that various kinds of large $N$ re-scalings of the couplings $R_{abcd}$ might lead to models where approximate fixed points could be found using only the two loop beta functions.  Since the requirement of conformal invariance only appears in the limit of semi-classical gravity, corresponding to large $N$, this might be sufficient. Again, we will have to leave the explorations of such speculations to future research.

\begin{center}
{\bf Acknowledgments }\\
\end{center} 
\hspace{6mm}S. A thanks D. Rogerson, R. Singh, and F. Stubbs for discussions. The authors thank S. Lukyanov and D. Diaconescu for helpful comments. The work of T. B. and S. A is supported in part by the DOE under grant DE-SC0010008.

\appendix\label{sec:app}
\section{Gamma Matrix Identities and Fierz Transformations in 2D}\label{sec:appendixA}
\hspace{6mm}In the correlation function calculations, we used the following gamma matrices identities in two dimensions, where the metric $g^{\mu\nu}=diag(+,-)$: 
\begin{equation}\label{eq:gammaidentities}
    \begin{split}
        \{\gamma^{\mu},\gamma^{\nu}\} &= 2g^{\mu\nu}\mathbb{I}_2 \\
        \gamma^{\mu}\gamma_{\mu} &= 2\mathbb{I}_2 \\
        \gamma^{\mu}\gamma^{\nu}\gamma_{\mu} &= 0 \\
        \gamma^{\mu}\gamma^3\gamma_{\mu} &= -2\gamma^3 \\
        \gamma^{\mu}\gamma^{\nu}\gamma^{\rho}\gamma_{\mu} &= 2\gamma^{\rho}\gamma^{\nu} \\
        \gamma^{\mu}(\gamma^{\nu_1}\dots\gamma^{\nu_{2k+1}})\gamma_{\mu}&=0 \\
        \Tr(\gamma^{\mu}\gamma^{\nu}) &= 2g^{\mu\nu} \\
        \Tr(\gamma^{\mu}\gamma^{\nu}\gamma^{\rho}\gamma^{\sigma}) &= 2(g^{\mu\nu}g^{\rho\sigma}-g^{\mu\rho}g^{\nu\sigma}+g^{\mu\sigma}g^{\nu\rho}) \\
        \gamma^{\mu}\gamma^{\nu}\gamma^{\rho} &= g^{\mu\nu}\gamma^{\rho}+g^{\nu\rho}\gamma^{\mu}-g^{\mu\rho}\gamma^{\nu} \\
        & = \gamma^{\rho}\gamma^{\nu}\gamma^{\mu} \\
        \gamma^{\mu}\gamma^{\nu}\gamma^{\rho}\gamma^{\sigma}\gamma^{\lambda} &= (g^{\mu\nu}g^{\rho\sigma}+g^{\nu\rho}g^{\mu\sigma}-g^{\mu\rho}g^{\nu\sigma})\gamma^{\lambda} \\
        & \hspace{5mm} -(g^{\mu\nu}g^{\rho\lambda}+g^{\nu\rho}g^{\mu\lambda} -g^{\mu\rho}g^{\nu\lambda})\gamma^{\sigma} \\
        &\hspace{10mm} + g^{\mu\nu}g^{\sigma\lambda}\gamma^{\rho}+g^{\nu\rho}g^{\sigma\lambda}\gamma^{\mu}-g^{\mu\rho}g^{\sigma\lambda}\gamma^{\nu} \\
        &= \gamma^{\lambda}\gamma^{\sigma}\gamma^{\rho}\gamma^{\nu}\gamma^{\mu} 
    \end{split}
\end{equation}

The current-current interaction is unique up to Fierz transformations, that is to say any four-fermi interaction can be rearranged to the current-current type using the Fierz transformation. The Fierz identity can be written as 
\begin{equation}
    \begin{split}
        (\overline{u}_1\Gamma^Au_2)(\overline{u}_3\Gamma^Bu_4) = \sum_{C,D}{C^{AB}}_{CD}(\overline{u}_1\Gamma^Du_4)(\overline{u}_3\Gamma^Du_2).
    \end{split}
\end{equation}
Where $\Gamma^A \in \{\mathbb{I}_2,\gamma^0,i\gamma^1,\gamma^3\}$ is any combinations of the normalized elements (with the convention $\Tr(\Gamma^A\Gamma^B)=2\delta^{AB}$).
The set in general dimensions is $\{\mathbb{I},\gamma^{\mu},\sigma^{\mu\nu}= \frac{i}{2}[\gamma^{\mu},\gamma^{\nu}], \gamma^3\gamma^{\mu},\gamma^3\}$. Left multiply $(\overline{u}_2\Gamma^Fu_3)(\overline{u}_4\Gamma^Eu_1)$, then
\begin{equation}
    \begin{split}
        LHS &= \overline{u}_4\Gamma^E\Gamma^A\Gamma^F\Gamma^Bu_4 = \Tr(\Gamma^E\Gamma^A\Gamma^F\Gamma^B) \\
        RHS &= \sum_{C,D}{C^{AB}}_{CD}\Tr(\Gamma^E\Gamma^C)\Tr(\Gamma^F\Gamma^D) = 4\sum_{C,D}{C^{AB}}_{CD}\delta^{EC}\delta^{FD} = 4{C^{AB}}_{EF} \\
        \Longrightarrow {C^{AB}}_{CD} &= \frac{1}{4}\Tr(\Gamma^C\Gamma^A\Gamma^D\Gamma^B)
    \end{split}
\end{equation}
\begin{equation}\label{eq:fierz}
    \begin{split}
        (\overline{u}_1\gamma^{\mu}u_2)(\overline{u}_3\gamma_{\mu}u_4) &= \sum_{C,D}\frac{1}{4}\Tr(\Gamma^C\gamma^{\mu}\Gamma^D\gamma_{\mu})(\overline{u}_1\Gamma^Cu_4)(\overline{u}_3\Gamma^Du_2) \\
        &= \sum_C\frac{1}{4}\Tr(\Gamma^C\gamma^{\mu}\Gamma^C\gamma_{\mu})(\overline{u}_1\Gamma^Cu_4)(\overline{u}_3\Gamma^Cu_2) \\
        &= \frac{1}{4}\Tr(\gamma^{\mu}\gamma_{\mu})(\overline{u}_1u_4)(\overline{u}_3u_2)+\frac{1}{4}\Tr(\gamma^3\gamma^{\mu}\gamma^3\gamma_{\mu})(\overline{u}_1\gamma^3u_4)(\overline{u}_3\gamma^3u_2) \\
        &=(\overline{u}_1u_4)(\overline{u}_3u_2)-(\overline{u}_1\gamma^3u_4)(\overline{u}_3\gamma^3u_2)
    \end{split}
\end{equation}

\begin{equation}\label{eq:fierz2}
    \begin{split}
        (\overline{u}_1u_2)(\overline{u}_3u_4)&= \sum_C\frac{1}{4}\Tr(\Gamma^C\Gamma^C)(\overline{u}_1\Gamma^Cu_4)(\overline{u}_3\Gamma^Cu_2) \\
        &= \frac{1}{2}(\overline{u}_1u_4)(\overline{u}_3u_2)+\frac{1}{2}(\overline{u}_1\gamma^0u_4)(\overline{u}_3\gamma^0u_2) \\
        &\hspace{10mm} -\frac{1}{2}(\overline{u}_1\gamma^1u_4)(\overline{u}_3\gamma^1u_2)+\frac{1}{2}(\overline{u}_1\gamma^3u_4)(\overline{u}_3\gamma^3u_2) \\
        &= \frac{1}{2}(\overline{u}_1u_4)(\overline{u}_3u_2) + \frac{1}{2}(\overline{u}_1\gamma^3u_4)(\overline{u}_3\gamma^3u_2) + \frac{1}{2}(\overline{u}_1\gamma^{\mu}u_4)(\overline{u}_3\gamma_{\mu}u_2) 
    \end{split}
\end{equation}
\begin{equation}\label{eq:fierz3}
    \begin{split}
        (\overline{u}_1\gamma^3u_2)(\overline{u}_3\gamma^3u_4) &= \sum_C\frac{1}{4}\Tr(\Gamma^C\gamma^3\Gamma^C\gamma^3)(\overline{u}_1\Gamma^Cu_4)(\overline{u}_3\Gamma^Cu_2) \\
        &= \frac{1}{4}\Tr(\gamma^3\gamma^3)(\overline{u}_1u_4)(\overline{u}_3u_2) + \frac{1}{4}\Tr((\gamma^0\gamma^3)^2)(\overline{u}_1\gamma^0u_4)(\overline{u}_3\gamma^0u_2) \\
        &\hspace{10mm} -\frac{1}{4}\Tr(-(\gamma^1\gamma^3)^2)(\overline{u}_1\gamma^1u_4)(\overline{u}_3\gamma^1u_2) - \frac{1}{4}\Tr(-(\gamma^3)^4)(\overline{u}_1\gamma^3u_4)(\overline{u}_3\gamma^3u_2) \\
        &= \frac{1}{2}(\overline{u}_1u_4)(\overline{u}_3u_2) -\frac{1}{2}(\overline{u}_1\gamma^0u_4)(\overline{u}_3\gamma^0u_2) \\
        &\hspace{10mm} +\frac{1}{2}(\overline{u}_1\gamma^1u_4)(\overline{u}_3\gamma^1u_2) + \frac{1}{2}(\overline{u}_1\gamma^3u_4)(\overline{u}_3\gamma^3u_2) \\
        &=\frac{1}{2}(\overline{u}_1u_4)(\overline{u}_3u_2) - \frac{1}{2}(\overline{u}_1\gamma^{\mu}u_4)(\overline{u}_3\gamma_{\mu}u_2) + \frac{1}{2}(\overline{u}_1\gamma^3u_4)(\overline{u}_3\gamma^3u_2) 
    \end{split}
\end{equation}

\section{Elimating Vanishing Diagrams in the Current-Current Correlation Function}\label{sec:appendixB}

\begin{table}
        \centering
        \begin{tabular}{c|c|c|c|}
        \cline{2-4}
            & & &  \\
            & 1 & $\exists$ \hspace{1mm} $n_i > 6$ &   $\gamma^{\mu} (\cdots) \gamma_{\mu} = 0$\\ 
            & & &  \\ \cline{2-4}
            & & &  \\
            \checkmark & 2 & 6+6 &  $\Tr(\cdots)\Tr(\cdots)$ \\
            & & &  \\ \cline{2-4}
            & & &  \\
            & 3 & 4+4+4 &  \begin{tabular}{c}
               ext(2+2+0) \\
               int(0+0+2)
            \end{tabular} $\overset{F.P.}{\Longrightarrow}$ \hspace{1mm} 0 \\
            & & &  \\ \cline{2-4}
            & & &  \\
            & 4 & $\exists$ \hspace{1mm} $n_i$ = 2 &  $\braket{\overline{\psi}\psi}$ correction \hspace{1mm} $\Longrightarrow$ 0  \\ 
            & & &  \\ \cline{2-4}
        \end{tabular}
        \caption{At one-loop level, 12 gamma matrices can be grouped to form different-sized loops, there are 4 different cases listed on the left. On the right, the reason why each case contributes or not is shown. Only the second case contributes to the current-current correlation function.}
        \label{table:1loop}
    \end{table}

\hspace{6mm}At the one-loop level, the $6!=720$ Wick contractions reduce to 10 distinct Feynman diagrams, which is still relatively a large number to read off and calculate the correlation functions. Luckily, observations made in the field strength correction calculation in Subsection \ref{subsec:1loop} become useful in determining which diagrams contribute and which vanish. 

The current-current correlation function at the one-loop level corresponds to the second-order expansion of the exponential in Eq.(\ref{eq:correlationfcn}) which consists of 6 currents, i.e.  $\overline{\psi}\gamma^{\mu}\psi$. The 6 currents contain 6 $\gamma$ matrices and 6 fermion-antifermion pairs in the function. Through Wick contraction, the 6 fermion-antifermion pairs give 6 fermionic propagators 
\begin{equation}
    \begin{split}
        D_F(x-y) \propto \int \frac{d^2p}{(2\pi)^2} \frac{i\slashed{p}}{p^2-m^2},
    \end{split}
\end{equation}
where the numerator $\slashed{p}=p_{\mu}\gamma^{\mu}$, so the 6 propagators contribute 6 more $\gamma$'s and in total there are 12 in the expression. The $\gamma$'s are traced over in several traces and the number of traces depends on how the fermionic fields are contracted. Any trace must contain an even number of $\gamma$'s because any propagator is connected to 2 $\gamma$'s and for a fermion loop consisting of $P\geq 1$ propagators, it is a trace over $P+2P/2=2P$ Dirac matrices, where the first $P$ comes from $\gamma$'s in the propagators and there are $2P$ $\gamma$'s connected to each propagator, divided by 2 due to double counting. Another way to see this is that any $\gamma$ provided by a propagator (propagator gamma) must be accompanied by a $\gamma$ from the current $\overline{\psi}\gamma^{\mu}\psi$ (current gamma), therefore any trace contains an even number of gamma matrices. Then determining which diagrams contribute becomes a simple math problem. Break the 12 $\gamma$'s into several sets, 
\begin{equation}
    12 = \sum_{i=1}^k n_i \hspace{3mm} \text{where} \hspace{3mm} 2 \leq n_i\leq 12 \hspace{3mm} \text{and} \hspace{3mm} n_i\in 2\mathbb{Z},
\end{equation}
where $k$ is the number of fermion loops (number of traces) in the diagram, and $n_i$ is the number of $\gamma$'s in a trace. Then we have the following cases:

1. $\exists$ $i$ such that $n_i > 6$: this case does NOT contribute because of $\gamma$-matrices identities in two dimensions. By the pigeonhole principle, any loop containing more than 3 current gammas must have $\gamma^{\mu}(\cdots)\gamma_{\mu}$ where $(\cdots)$ are an odd number of gamma matrices, thus it vanishes for the same reason shown in Eq.(\ref{eq:2pt1loopdiag1}).

2. $k=2$ and $n_i=6$ for $\forall$ $i$: this case gives diagrams consisting of two equal loops of the same size, which contribute to the current-current correlation function.

3. $k=3$ and $n_i=4$ for $\forall$ $i$: this case does NOT contribute. One of the three loops contains both incoming momentum flows; one contains both outgoing momentum flows; one contains 2 internal propagators with momentum flows $p, p'$. We denote such configurations as

\begin{table}[H]
    \centering
    \begin{tabular}{c}
        ext(2+2+0) \\
        int(0+0+2)
    \end{tabular}
    \label{tab:my_label}
\end{table}
\noindent where "ext" and "int" stand for "external" and "internal" momentum flow, and each column represents a loop containing a number of external and internal momentum flows. We can easily see that in the third loop, $p'$ linearly depends on the undetermined momentum $p$, i.e. $p'=p'(p)\propto p $, and when performing Feynman parametrization, the third loop contributes a term that is 
\begin{equation}\label{eq:fp0}
    \begin{split}
        \sim \Tr\big(\slashed{p}\gamma^{\mu}\slashed{p}'\gamma^{\nu}\big) = p_{\alpha}p'_{\beta}\Tr\big(\gamma^{\alpha}\gamma^{\mu}\gamma^{\beta}\gamma^{\nu}\big) \sim  g_{\alpha\beta}p^2\Tr\big(\gamma^{\alpha}\gamma^{\mu}\gamma^{\beta}\gamma^{\nu}\big) = p^2 \Tr\big(\gamma_{\beta}\gamma^{\mu}\gamma^{\beta}\gamma^{\nu}\big) =0,
    \end{split}
\end{equation}
as shown in Eq.(\ref{eq:feynmanpara}).

4. $\exists$ $i$ such that $n_i = 2$: this case does NOT contribute because such diagrams either are disconnected or vanish for the same reason shown in Eq.(\ref{eq:2pt1loopdiag2}) due to integration of an odd function.

Table.\ref{table:1loop} is a summary of the 4 cases and their reasons for whether or not they contribute.

At the two-loop level, the $8!=40320$ Wick contractions reduce to $96$ distinct Feynman diagrams. Applying the same reasoning above, $90$ of which vanishes and only $6$ diagrams remain. Group $16$ gamma matrices into loops and we have $6$ cases. Table.\ref{table:2loop} gives a summary of the 6 cases. Cases $1,2,5$, and $6$ follow the same reasons as cases $1,2,3$, and $4$, respectively, at the one-loop level. Cases $3$ and $4$ are more complicated which we will explain in the following. 

\begin{table}
        \centering
        \begin{tabular}{c|c|c|c|}
        \cline{2-4}
            & & & \\
            & 1 & $\exists$ \hspace{1mm} $n_i > 8$ &  $\gamma^{\mu} (\cdots) \gamma_{\mu} = 0$\\
            & & & \\ \cline{2-4}
            & & &  \\
            \checkmark & 2 & 8+8 &  $\Tr(\cdots)\Tr(\cdots)$ \\
            & & & \\ \cline{2-4}
            & & &  \\
            & 3 & 8+4+4 &  \begin{tabular}{c c c}
               ext(0+2+2) & ext(2+2+0) & ext(4+0+0) \\
               int(4+0+0) & int(2+0+2) & int(0+2+2)
            \end{tabular} $\overset{F.P.}{\Longrightarrow}$ \hspace{1mm} 0 \\
            & & & \\ \cline{2-4}
            & & &  \\
            & 4 & 6+6+4 &  \begin{tabular}{c c}
               ext(0+2+2) & ext(2+2+0) \\
               int(3+1+0) & int(1+1+2)
            \end{tabular} $\overset{F.P.}{\Longrightarrow}$ \hspace{1mm} 0 \\
            & & & \\ \cline{2-4}
            & & &  \\
            & 5 & 4+4+4+4 &  \begin{tabular}{c}
               ext(2+2+0+0) \\
               int(0+0+2+2)
            \end{tabular} $\overset{F.P.}{\Longrightarrow}$ \hspace{1mm} 0 \\
            & & & \\ \cline{2-4}
            & & &  \\
            & 6 & $\exists$ \hspace{1mm} $n_i$ = 2 &  $\braket{\overline{\psi}\psi}$ correction \hspace{1mm} $\Longrightarrow$ \hspace{1mm} 0  \\ 
            & & & \\ \cline{2-4}
        \end{tabular}
        \caption{A summary of whether or not a configuration contributes at the two-loop level, there are $6$ ways to group the $16$ gamma matrices and only the second case contributes. Cases $3$ and $4$ have $3$ and $2$ subcases, respectively.}
        \label{table:2loop}
    \end{table}
For case $3$, there are $3$ ways to form $3$ loops consisting $8, 4$, and $4$ gamma matrices, shown in Table.\ref{table:2loop} case 3:

3.1) One loop contains $4$ internal flows and two loops contain $2$ external flows each. The loop with $4$ internal flows $p,p',q,q'$ yields 0 by Feynman parametrization, as $p'=p'(p)$ and $q'=q'(q)$ similar to the case in Eq.(\ref{eq:fp0}).

3.2) One loop contains $2$ external and $2$ internal flows, one contains $2$ external flows, and one contains $2$ internal flow. The loop with only $2$ external flows forces one of the other two loops to contain a pair of linearly dependent internal flows, and such diagrams vanish similar to the case in Eq.(\ref{eq:fp0}).

3.3) One loop contains $4$ external flows, and two loops each contain $2$ internal flows. The loop with $4$ external flows is in $\Tr[\gamma^{\mu}(\cdots)\gamma_{\mu}\cdots]$ form, and it vanishes by two-dimensional gamma matrices identities.

For case 4, $2$ ways to form $3$ loops both vanish:

4.1) One loop contains $3$ internal flows, one contains $2$ external flows, and one contains $2$ external and $1$ internal flows. Similar to case 3.1), the loop with only $2$ external flows forces one of the other two loops to contain a pair of linearly dependent internal flows, thus this case vanishes.

4.2) Two loops contain $2$ external and $1$ internal flows each, and one loop contains $2$ internal flows. In this case, one can verify that $3$ out of the $4$ internal flows depend on one of the two undetermined momenta, thus the integrands are odd functions of undetermined momenta. Such diagrams also vanish.

Therefore, the only cases left at the two-loop level are diagrams consisting of $2$ equal-sized loops of $8$ gamma matrices.

\section{Color maps of solutions for Majorana fermions and odd behaviour for $N=4$}\label{sec:appendixD}
\hspace{6mm}We present color maps for Majorana fermions in this appendix and explain behaviours of $N=4$ for Majorana cases.

\begin{figure}[H]
\begin{minipage}{0.33\textwidth}
    \centering
    \includegraphics[scale = 0.46]{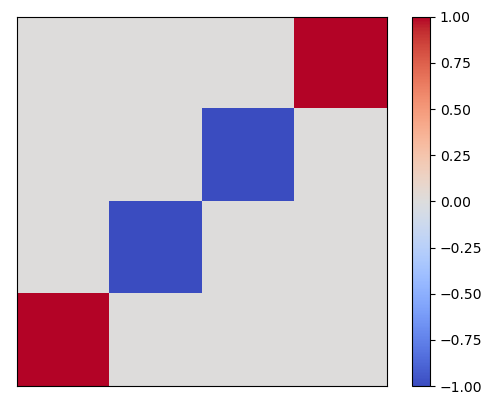}
    \caption{N=2 Majorana.}
    \label{fig:2maj}
\end{minipage}
\begin{minipage}{0.33\textwidth}
    \centering
    \includegraphics[scale = 0.46]{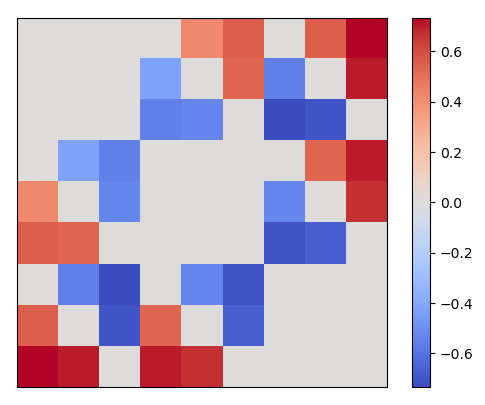}
    \caption{N=3 Majorana.}
    \label{fig:3maj}
\end{minipage}
\begin{minipage}{0.33\textwidth}
    \centering
    \includegraphics[scale = 0.46]{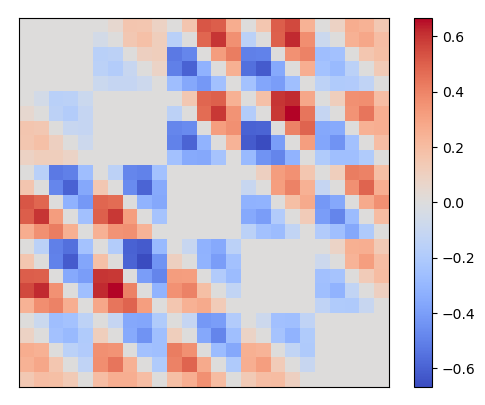}
    \caption{N=5 Majorana.}
    \label{fig:5maj}
\end{minipage}
\end{figure}

\begin{figure}[H]
\begin{minipage}{0.33\textwidth}
    \centering
    \includegraphics[scale = 0.46]{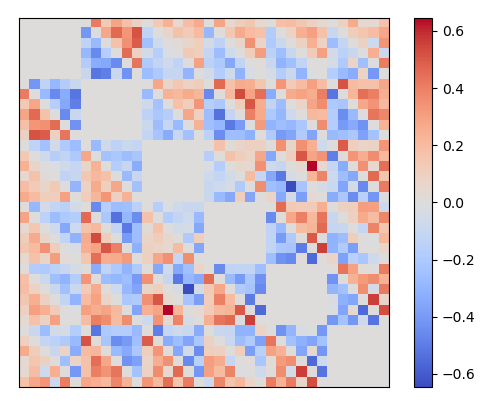}
    \caption{N=6 Majorana.}
    \label{fig:6maj}
\end{minipage}
\begin{minipage}{0.33\textwidth}
    \centering
    \includegraphics[scale = 0.46]{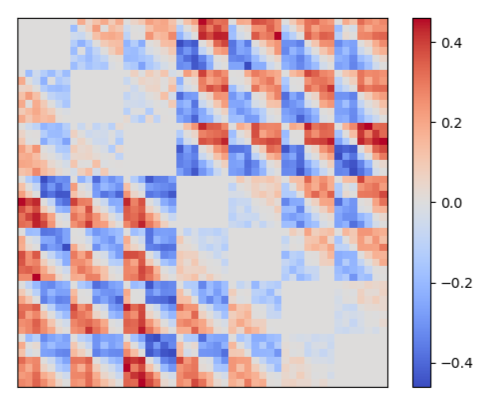}
    \caption{N=7 Majorana.}
    \label{fig:7maj}
\end{minipage}
\begin{minipage}{0.33\textwidth}
    \centering
    \includegraphics[scale = 0.46]{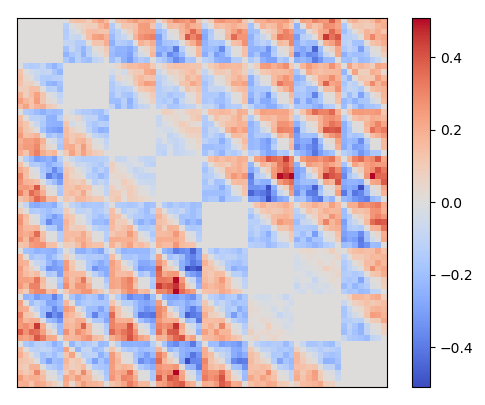}
    \caption{N=8 Majorana.}
    \label{fig:8maj}
\end{minipage}
\end{figure}

The coloring procedure for Majorana is the same as the one for Dirac, we first arrange all couplings into a matrix of matrices shown in Eq.(\ref{eq:matrix}). The antisymmetry properties in the first pairs of indices lead to diagonal symmetry in the big matrix and the antisymmetry properties in the second pairs of indices lead to diagonal symmetry in the submatrices. The antisymmetry in indices gives vanishing diagonal terms. All independent couplings are covered by these upper triangle parts, with $R_{abab}$'s appearing only once and all the rest nonvanishing couplings appearing twice. For example, $N=3$ has $6$ DOFs, where $R_{1213}, R_{1223},R_{1323}$ repeat themselves, thus the $9$ red blocks at the top right corner. As one can see in Figures.\ref{fig:2maj}-\ref{fig:8maj}, all allowed couplings are turned on in a solution. In Majorana cases, no obvious pattern is observed other than the antisymmetry and symmetry properties in the indices, unlike in Dirac cases where one similar pattern of solution emerges for all $N$.

\begin{figure}[H]
\begin{minipage}{0.33\textwidth}
    \centering
    \includegraphics[scale = 0.46]{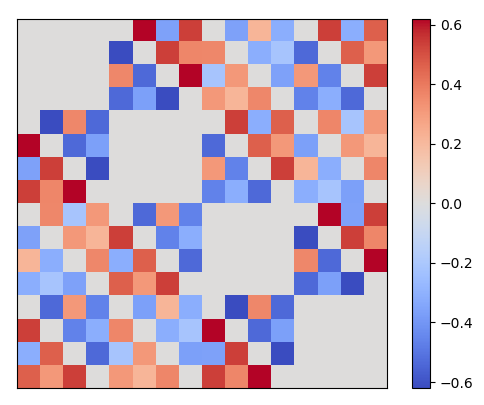}
\end{minipage}
\begin{minipage}{0.33\textwidth}
    \centering
    \includegraphics[scale = 0.46]{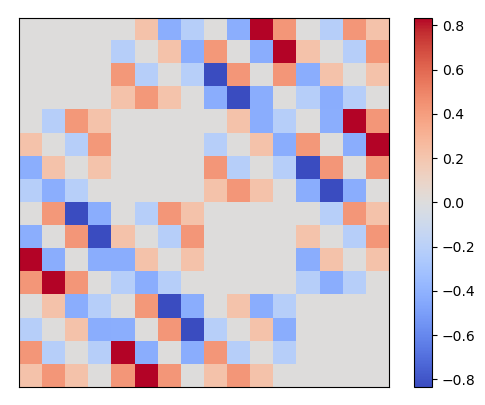}
\end{minipage}
\begin{minipage}{0.33\textwidth}
    \centering
    \includegraphics[scale = 0.46]{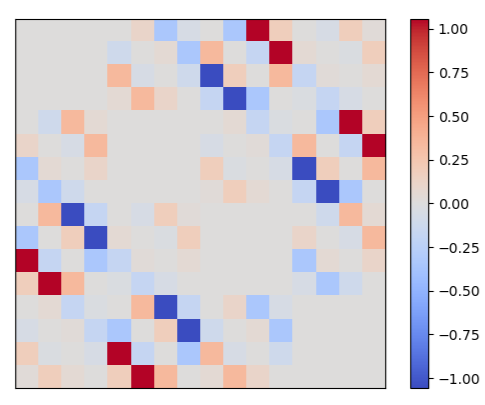}
\end{minipage}
    \caption{The color map on the left is an example of a general solution to the vanishing beta functions for $N=4$ Majorana fermions. The middle color map is a solution where couplings become weaker except for $R_{1313}, R_{2424}$, and $R_{1324}$. The concentration continues in the color map on the right.}
    \label{fig:4majgeneral}
\end{figure}

\begin{figure}[H]
\begin{minipage}{0.33\textwidth}
    \centering
    \includegraphics[scale = 0.46]{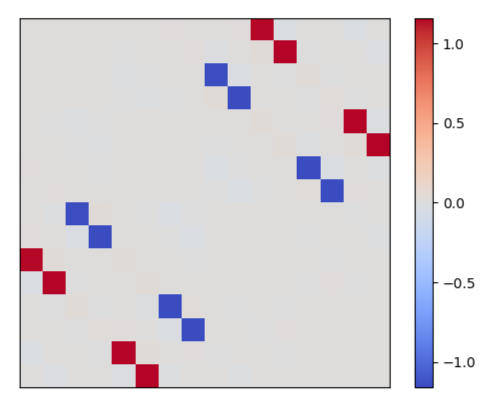}
\end{minipage}
\begin{minipage}{0.33\textwidth}
    \centering
    \includegraphics[scale = 0.46]{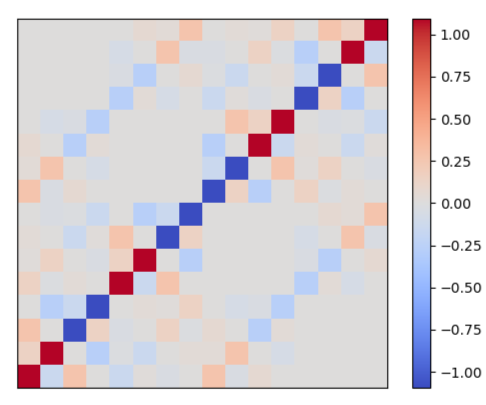}
\end{minipage}
\begin{minipage}{0.33\textwidth}
    \centering
    \includegraphics[scale = 0.46]{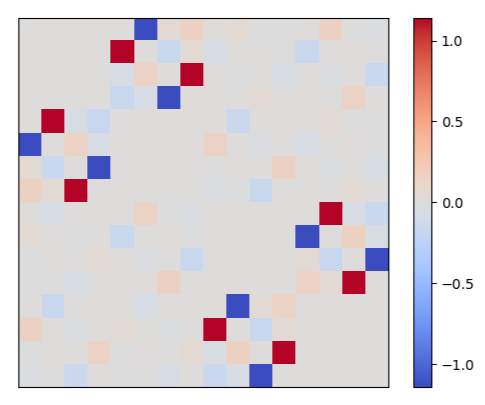}
\end{minipage}
\caption{Three different ways of coupling concentration. The couplings concentrate on $R_{1313}, R_{2424}$, and $R_{1324}$ in the diagram on the left. The couplings concentrate on $R_{1414}, R_{2323}$, and $R_{1423}$ in the middle diagram. The couplings concentrate on $R_{1212}, R_{3434}$, and $R_{1234}$ in the diagram on the right.}
\label{fig:4maj3ways}
\end{figure}

\begin{figure}[H]
\begin{minipage}{0.33\textwidth}
    \centering
    \includegraphics[scale = 0.46]{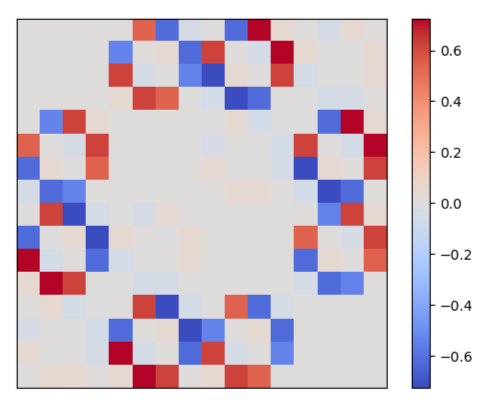}
\end{minipage}
\begin{minipage}{0.33\textwidth}
    \centering
    \includegraphics[scale = 0.46]{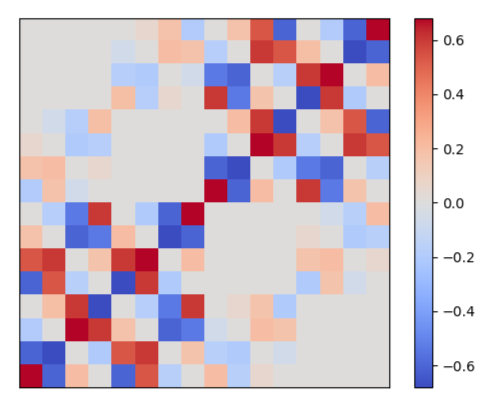}
\end{minipage}
\begin{minipage}{0.33\textwidth}
    \centering
    \includegraphics[scale = 0.46]{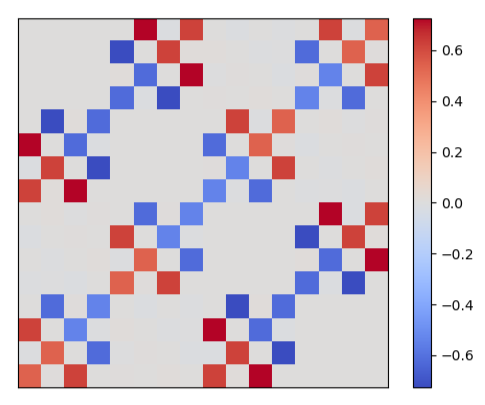}
\end{minipage}
    \caption{Three patterns emerge when $10$ DOFs survive the coupling concentration. These solutions follow the form in Eq.(\ref{eq:soln}). Left: $R_{1212},R_{3434}, R_{1313}, R_{2424}$ and their cross couplings. Middle: $R_{1313}, R_{2424}, R_{1414}, R_{2323}$ and their cross couplings. Right: $R_{1212}, R_{3434}, R_{1414}, R_{2323}$ and their cross couplings.}
    \label{fig:4maj10dof}
\end{figure}

However, color maps show different 
patterns for $N=4$ Majorana fermions shown in Figures.\ref{fig:4majgeneral}-\ref{fig:4maj10dof} for different numerical solutions that generate different patterns. Figure.\ref{fig:4majgeneral} gives an example of how couplings concentrate more and more and eventually settle at $3$ DOFs $R_{1313}, R_{2424}$, and cross coupling $R_{1324}$. The cross coupling of $R_{abab}$ and $R_{cdcd}$ is defined as $R_{abcd}$. Similarly, couplings can also settle at $R_{1414}, R_{2323}$, and $R_{1423}$ as shown in the middle color map in Figure.\ref{fig:4maj3ways}, or $R_{1212}, R_{3434}$, and $R_{1234}$ as shown in the color map on the right in Figure.\ref{fig:4maj3ways}. These are the only three patterns with $3$ DOFs, and the form of Eq.(\ref{eq:soln}) is satisfied by these solutions. A theory with this structure is equivalent to a theory of two bosons with an interacting term in the bosonization picture where a free boson is equivalent to two real fermions with a four-fermion interaction. 

Other middle stages of coupling concentration are shown in Figure.\ref{fig:4maj10dof}. These are the only patterns with $10$ DOFs. Once couplings completely concentrate following these patterns, these stages are solutions in the form of Eq.(\ref{eq:soln}). In the left most color map in Figure.\ref{fig:4maj10dof}, couplings concentrate on $R_{1313}, R_{2424}, R_{1414}, R_{2323}$ and their cross couplings $R_{1324}, R_{1314}, R_{1323}, R_{1423}, R_{1424}, R_{2324}$. The middle one in Figure.\ref{fig:4maj10dof} is an example of couplings concentrating on $R_{1212}, R_{3434}, R_{1313}, R_{2424}$ and their 6 cross couplings. The color map on the right in Figure.\ref{fig:4maj10dof} shows couplings concentrate on $R_{1212}, R_{3434}, R_{1414}, R_{2323}$ and their 6 cross couplings. A further step is to explore whether the fermionic theory with such solutions has a bosonic equivalence, if this is the case, is there a way to map the four-fermion couplings to couplings of this bosonic theory? 

For $N>4$ Majorana fermions, patterns of solutions become more and more difficult to observe as $N$ gets larger, therefore, further work is needed to determine the most general symbolic solutions to the vanishing beta functions up to the two-loop level.

\newpage
\section{Samples of Numerical Results to Vanishing Beta Functions}\label{sec:appendixC}

\begin{table}[H]
\begin{center}
    \begin{tabular}{|c|c|}
        \hline
        $\begin{matrix}R_{1111} & R_{1112}  \\
        R_{1121} & R_{1122} \end{matrix}$ & $\begin{matrix}R_{1112} & R_{1212} \\
        R_{1221} & R_{1222} \end{matrix}$ \\
        \hline
        $\begin{matrix}R_{1121} & R_{1221} \\
        R_{2121} & R_{2122} \end{matrix}$ & $\begin{matrix}R_{1122} & R_{1222} \\
        R_{2122} & R_{2222} \end{matrix}$ \\
        \hline
    \end{tabular}
    \begin{tabular}{|c c|c c|}
        \hline
        -4.3860748e-2 &  -2.4213040e-1 & -2.4213040e-1 &  3.1556370e-2 \\
        3.0402219e-1 & 2.4040258e-1 & 6.2101346e-3 & -2.4448646e-1 \\
        \hline
        3.0402219e-1 &  6.2101346e-3 & 2.4040258e-1 & -2.4448646e-1 \\
        1.2221232e-3 & 3.0355853e-1 & 3.0355853e-1 &  5.2484182e-1 \\
        \hline
    \end{tabular} 
    \caption{A solution to the vanishing beta functions for the theory with N=2 Dirac fermions that does not follow the form of Eq.(\ref{eq:soln}).}
    \label{table:2dirac} 
\end{center}
\end{table}

\begin{table}[H]
\centering
    \begin{tabular}{|c|}
        \hline
        $\begin{matrix}R_{1111} & R_{1112} & R_{1113} \\R_{1121} & R_{1122} & R_{1123}\\ R_{1131} & R_{1132} & R_{1133}\end{matrix}$ \\
        \hline
        $\begin{matrix}R_{1112} & R_{1212} & R_{1213}\\R_{1221} & R_{1222} & R_{1223}\\R_{1231} & R_{1232} & R_{1233}\end{matrix}$ \\
        \hline
        $\begin{matrix}R_{1113} & R_{1213} & R_{1313}\\R_{1321} & R_{1322} & R_{1323}\\R_{1331} & R_{1332} & R_{1333}\end{matrix}$ \\
        \hline
        $\begin{matrix}R_{1121} & R_{1221} & R_{1321}\\R_{2121} & R_{2122} & R_{2123}\\R_{2131} & R_{2132} & R_{2133}\end{matrix}$\\
        \hline
        $\begin{matrix}R_{1122} & R_{1222} & R_{1322}\\R_{2122} & R_{2222} & R_{2223}\\R_{2231} & R_{2232} & R_{2233}\end{matrix}$\\
        \hline
        $\begin{matrix}R_{1123} & R_{1223} & R_{1323}\\R_{2123} & R_{2223} & R_{2323}\\R_{2331} & R_{2332} & R_{2333}\end{matrix}$\\
        \hline
        $\begin{matrix}R_{1131} & R_{1231} & R_{1331}\\R_{2131} & R_{2231} & R_{2331}\\R_{3131} & R_{3132} & R_{3133}\end{matrix}$\\
        \hline
        $\begin{matrix}R_{1132} & R_{1232} & R_{1332}\\R_{2132} & R_{2232} & R_{2332}\\R_{3132} & R_{3232} & R_{3233}\end{matrix}$\\
        \hline
        $\begin{matrix}R_{1133} & R_{1233} & R_{1333}\\R_{2133} & R_{2233} & R_{2333}\\R_{3133} & R_{3233} & R_{3333}\end{matrix}$\\
        \hline
    \end{tabular}
    \begin{tabular}{|c c c|}
        \hline
        5.87648556482932e-1 & 4.15490588387255e-1 & 8.09728843069292e-2\\-1.17589376850528e-1 & 6.35939476084822e-1 & 9.66442479189108e-2\\-1.09541107265992e-1 & -2.52992302722233e-1 & 8.71030751509361e-2 \\
        \hline 
        4.15490588387255e-1 & 5.59563619288908e-4 & 3.28299313172747e-4\\5.37601466194129e-4 & 4.1563647157943e-1 & 3.72746485831559e-4\\4.94452190544431e-4 & 4.14961112540219e-4 & 4.15297389252339e-1 \\
        \hline
        8.09728843069292e-2 & 3.28299313172747e-4 & 2.08068902716359e-4\\3.52770559599846e-4 & 8.10293435524669e-2 & 2.32760156206622e-4\\2.74020612749278e-4 & 2.8670364435102e-4 & 8.08365134052374e-2 \\
        \hline
        -1.17589376850528e-1 & 5.37601466194129e-4 & 3.52770559599846e-4\\6.28702609173722e-4 & -1.1754154676859e-1 & 3.95675644307164e-4\\4.52106341843372e-4 & 4.91213947244407e-4 & -1.17842968195166e-1 \\
        \hline
        6.35939476084822e-1 & 4.1563647157943e-1 & 8.10293435524669e-2\\-1.1754154676859e-1 & 6.84345214578727e-1 & 9.67115256538758e-2\\-1.09396141048504e-1 & -2.52955300347104e-1 & 1.35399551371414e-1\\
        \hline
        9.66442479189108e-2 & 3.72746485831559e-4 & 2.32760156206622e-4\\3.95675644307164e-4 & 9.67115256538758e-2 & 2.61029933264969e-4\\3.10949258252192e-4 & 3.20139640173062e-4 & 9.64918473705521e-2\\
        \hline
        -1.09541107265992e-1 & 4.94452190544431e-4 & 2.74020612749278e-4\\4.52106341843372e-4 & -1.09396141048504e-1 & 3.10949258252192e-4\\3.09068165743291e-4 & 4.59802986817494e-4 & -1.09686644813478e-1\\
        \hline
        -2.52992302722233e-1 & 4.14961112540219e-4 & 2.8670364435102e-4\\4.91213947244407e-4 & -2.52955300347104e-1 & 3.20139640173062e-4\\4.59802986817494e-4 & 2.9698568041399e-4 & -2.53201916372555e-1\\
        \hline
        8.71030751509361e-2 & 4.15297389252339e-1 & 8.08365134052374e-2\\-1.17842968195166e-1 & 1.35399551371414e-1 & 9.64918473705521e-2\\-1.09686644813478e-1 & -2.53201916372555e-1 & -4.13338584898267e-1\\
        \hline
    \end{tabular} 
    \caption{A solution for N=3 Dirac fermions.}
    \label{table:3dirac}
\end{table}

\begin{sidewaystable}
    \centering
    \begin{tabular}{|c|}
        \hline
        $\begin{matrix}R_{1111} & R_{1112} & R_{1113} & R_{1114}\\R_{1121} & R_{1122} & R_{1123} & R_{1124}\\R_{1131} & R_{1132} & R_{1133} & R_{1134}\\R_{1141} & R_{1142} & R_{1143} & R_{1144}\end{matrix}$ \\
        \hline
        $\begin{matrix}R_{1112} & R_{1212} & R_{1213} & R_{1214}\\R_{1221} & R_{1222} & R_{1223} & R_{1224}\\R_{1231} & R_{1232} & R_{1233} & R_{1234}\\R_{1241} & R_{1242} & R_{1243} & R_{1244}\end{matrix}$  \\
        \hline
        $\begin{matrix}R_{1113} & R_{1213} & R_{1313} & R_{1314}\\R_{1321} & R_{1322} & R_{1323} & R_{1324}\\R_{1331} & R_{1332} & R_{1333} & R_{1334}\\R_{1341} & R_{1342} & R_{1343} & R_{1344}\end{matrix}$  \\
        \hline
        $\begin{matrix}R_{1114} & R_{1214} & R_{1314} & R_{1414}\\R_{1421} & R_{1422} & R_{1423} & R_{1424}\\R_{1431} & R_{1432} & R_{1433} & R_{1434}\\R_{1441} & R_{1442} & R_{1443} & R_{1444}\end{matrix}$ \\
        \hline
        $\begin{matrix}R_{1121} & R_{1221} & R_{1321} & R_{1421}\\R_{2121} & R_{2122} & R_{2123} & R_{2124}\\R_{2131} & R_{2132} & R_{2133} & R_{2134}\\R_{2141} & R_{2142} & R_{2143} & R_{2144}\end{matrix}$  \\
        \hline
        $\begin{matrix}R_{1122} & R_{1222} & R_{1322} & R_{1422}\\R_{2122} & R_{2222} & R_{2223} & R_{2224}\\R_{2231} & R_{2232} & R_{2233} & R_{2234}\\R_{2241} & R_{2242} & R_{2243} & R_{2244}\end{matrix}$  \\
        \hline
        $\begin{matrix}R_{1123} & R_{1223} & R_{1323} & R_{1423}\\R_{2123} & R_{2223} & R_{2323} & R_{2324}\\R_{2331} & R_{2332} & R_{2333} & R_{2334}\\R_{2341} & R_{2342} & R_{2343} & R_{2344}\end{matrix}$  \\
        \hline
        $\begin{matrix}R_{1124} & R_{1224} & R_{1324} & R_{1424}\\R_{2124} & R_{2224} & R_{2324} & R_{2424}\\R_{2431} & R_{2432} & R_{2433} & R_{2434}\\R_{2441} & R_{2442} & R_{2443} & R_{2444}\end{matrix}$ \\
        \hline
    \end{tabular}
    \begin{tabular}{|c c c c|}
        \hline
        8.93891334289414e-2 & -9.08200765529855e-2 & 4.87993833917034e-2 & 2.23581578500429e-1\\1.73381012395297e-1 & -4.17010032954388e-2 & 4.6646543694454e-1 & 9.93087222968853e-2\\1.58558911566139e-1 & 4.52465020090685e-1 & 2.8389663637554e-2 & -1.63924330125004e-1\\ 4.75721783983157e-3 & -3.58994878445356e-1 & 1.12327371281389e-1 & 4.44100160950618e-1 \\
        \hline 
        -9.08200765529855e-2 &  1.71052767932729e-3 &  1.93060133709441e-3 &  1.89843302179749e-3\\ 2.96189119224102e-3 & -8.92895768322059e-2 &  1.97390000514186e-3 &  2.61074796896732e-3\\ 2.40165860870197e-3 &  2.56284197024156e-3 & -9.03723841098154e-2 &  2.11490288448325e-3\\ 2.35821991735503e-3 &  2.58807745278666e-3 &  1.8085886792796e-3 & -9.0800486359443e-2 \\
        \hline
        4.87993833917034e-2 &  1.93060133709441e-3 &  1.6936758398197e-3 &  2.03107883583002e-3\\ 1.54367500346283e-3 & 4.84899996107517e-2 &  1.86979604656575e-3 &  1.79080549133709e-3\\ 1.78430153581487e-3 &  1.92339007103261e-3 & 4.85049584020254e-2 &  1.43254060620137e-3\\ 1.25052863228629e-3 &  1.32845068929825e-3 &  1.07201615064092e-3 & 4.81159771724384e-2 \\
        \hline
        2.23581578500429e-1 &  1.89843302179749e-3 &  2.03107883583002e-3 &  1.47610977973788e-3\\ 2.0059157954446e-3 & 2.23846997738574e-1 &  2.18802237204685e-3 &  2.07659959472313e-3\\ 1.84087370176185e-3 &  1.90974014151239e-3 & 2.23370066864299e-1 &  2.08605758348406e-3\\ 1.7917763064753e-3 &  1.96278188235845e-3 &  1.30812316112946e-3 & 2.23151197796952e-1 \\
        \hline
        1.73381012395297e-1 &  2.96189119224102e-3 &  1.54367500346283e-3 &  2.0059157954446e-3\\ 2.82846144695018e-3 & 1.73671827323392e-1 &  2.98753320525608e-3 &  2.63791972200659e-3\\ 2.52104369314041e-3 &  2.63789857454227e-3 & 1.733136299339e-1 &  2.38810631084216e-3\\ 2.07794831327891e-3 &  2.12047093908491e-3 &  2.04441432655092e-3 & 1.73175284635749e-1 \\
        \hline
        -4.17010032954388e-2 & -8.92895768322059e-2 & 4.84899996107517e-2 & 2.23846997738574e-1\\1.73671827323392e-1 & -1.73818121724867e-1 & 4.67814084997426e-1 &  9.97096618974108e-2\\1.58987547899646e-1 & 4.52863700838672e-1 & -1.03119353024888e-1 & -1.63372737242689e-1\\ 4.78464959400363e-3 & -3.59139073909566e-1 & 1.12837363644479e-1 & 3.12850716672893e-1 \\
        \hline
        4.6646543694454e-1 &  1.97390000514186e-3 &  1.86979604656575e-3 &  2.18802237204685e-3\\ 2.98753320525608e-3 & 4.67814084997426e-1 &  1.92526733017194e-3 &  2.47974227969789e-3\\ 2.44946384671077e-3 &  2.60411172815349e-3 & 4.66717323041201e-1 &  2.23656869548097e-3\\ 2.28777290938034e-3 &  2.4482832432676e-3 &  2.3553065012906e-3 & 4.66282825250818e-1 \\
        \hline
         9.93087222968853e-2 &  2.61074796896732e-3 &  1.79080549133709e-3 &  2.07659959472313e-3\\ 2.63791972200659e-3 &  9.97096618974108e-2 &  2.47974227969789e-3 &  2.24391953895858e-3\\ 2.39419476780728e-3 &  2.5393522972734e-3 &  9.92881322122194e-2 &  2.05063520505214e-3\\ 1.88218275451014e-3 &  1.92510275257336e-3 &  1.63169569765752e-3 &  9.92424706998542e-2 \\
        \hline
    \end{tabular} 
    \caption{First half of a solution for N=4 Dirac fermions.}
    \label{table:4dirac1}
\end{sidewaystable}

\begin{sidewaystable}
    \centering
    \begin{tabular}{|c|}
        \hline
        $\begin{matrix}R_{1131} & R_{1231} & R_{1331} & R_{1431}\\R_{2131} & R_{2231} & R_{2331} & R_{2431}\\R_{3131} & R_{3132} & R_{3133} & R_{3134}\\R_{3141} & R_{3142} & R_{3143} & R_{3144}\end{matrix}$  \\
        \hline
        $\begin{matrix}R_{1132} & R_{1232} & R_{1332} & R_{1432}\\R_{2132} & R_{2232} & R_{2332} & R_{2432}\\R_{3132} & R_{3232} & R_{3233} & R_{3234}\\R_{3241} & R_{3242} & R_{3243} & R_{3244}\end{matrix}$  \\
        \hline
        $\begin{matrix}R_{1133} & R_{1233} & R_{1333} & R_{1433}\\R_{2133} & R_{2233} & R_{2333} & R_{2433}\\R_{3133} & R_{3233} & R_{3333} & R_{3334}\\R_{3341} & R_{3342} & R_{3343} & R_{3344}\end{matrix}$  \\
        \hline
        $\begin{matrix}R_{1134} & R_{1234} & R_{1334} & R_{1434}\\R_{2134} & R_{2234} & R_{2334} & R_{2434}\\R_{3134} & R_{3234} & R_{3334} & R_{3434}\\R_{3441} & R_{3442} & R_{3443} & R_{3444}\end{matrix}$ \\
        \hline
        $\begin{matrix}R_{1141} & R_{1241} & R_{1341} & R_{1441}\\R_{2141} & R_{2241} & R_{2341} & R_{2441}\\R_{3141} & R_{3241} & R_{3341} & R_{3441}\\R_{4141} & R_{4142} & R_{4143} & R_{4144}\end{matrix}$  \\
        \hline
        $\begin{matrix}R_{1142} & R_{1242} & R_{1342} & R_{1442}\\R_{2142} & R_{2242} & R_{2342} & R_{2442}\\R_{3142} & R_{3242} & R_{3342} & R_{3442}\\R_{4142} & R_{4242} & R_{4243} & R_{4244}\end{matrix}$  \\
        \hline
        $\begin{matrix}R_{1143} & R_{1243} & R_{1343} & R_{1443}\\R_{2143} & R_{2243} & R_{2343} & R_{2443}\\R_{3143} & R_{3243} & R_{3343} & R_{3443}\\R_{4143} & R_{4243} & R_{4343} & R_{4344}\end{matrix}$  \\ 
        \hline
        $\begin{matrix}R_{1144} & R_{1244} & R_{1344} & R_{1444}\\R_{2144} & R_{2244} & R_{2344} & R_{2444}\\R_{3144} & R_{3244} & R_{3344} & R_{3444}\\R_{4144} & R_{4244} & R_{4344} & R_{4444}\end{matrix}$ \\
        \hline
    \end{tabular}
    \begin{tabular}{|c c c c|}
        \hline
        1.58558911566139e-1 &  2.40165860870197e-3 &  1.78430153581487e-3 &  1.84087370176185e-3\\ 2.52104369314041e-3 & 1.58987547899646e-1 &  2.44946384671077e-3 &  2.39419476780728e-3\\ 2.27817944586346e-3 &  2.39833248770982e-3 & 1.5852468439324e-1 &  2.15416233905997e-3\\ 1.93903547033657e-3 &  2.03385597278184e-3 &  1.72377197400074e-3 & 1.58324230374422e-1 \\
        \hline 
        4.52465020090685e-1 &  2.56284197024156e-3 &  1.92339007103261e-3 &  1.90974014151239e-3\\ 2.63789857454227e-3 & 4.52863700838672e-1 &  2.60411172815349e-3 &  2.5393522972734e-3\\ 2.39833248770982e-3 &  2.52237722419909e-3 & 4.52406482033699e-1 &  2.28354266360319e-3\\ 2.02819472846199e-3 &  2.1239967070906e-3 &  1.79094217772047e-3 & 4.52211803714519e-1 \\
        \hline
        2.8389663637554e-2 & -9.03723841098154e-2 & 4.85049584020254e-2 & 2.23370066864299e-1\\1.733136299339e-1 & -1.03119353024888e-1 & 4.66717323041201e-1 &  9.92881322122194e-2\\1.5852468439324e-1 & 4.52406482033699e-1 & -3.30102173992401e-2 & -1.63742580345448e-1\\ 4.65639058481899e-3 & -3.59168172715776e-1 & 1.12426473698445e-1 & 3.83087886565564e-1 \\
        \hline
        -1.63924330125004e-1 &  2.11490288448325e-3 &  1.43254060620137e-3 &  2.08605758348406e-3\\ 2.38810631084216e-3 & -1.63372737242689e-1 &  2.23656869548097e-3 &  2.05063520505214e-3\\ 2.15416233905997e-3 &  2.28354266360319e-3 & -1.63742580345448e-1 &  1.83875902890197e-3\\ 1.81220258635743e-3 &  1.90480511326621e-3 &  1.75603803032357e-3 & -1.64191775784397e-1 \\
        \hline
         4.75721783983157e-3 &  2.35821991735503e-3 &  1.25052863228629e-3 &  1.7917763064753e-3\\ 2.07794831327891e-3 &  4.78464959400363e-3 &  2.28777290938034e-3 &  1.88218275451014e-3\\ 1.93903547033657e-3 &  2.02819472846199e-3 &  4.65639058481899e-3 &  1.81220258635743e-3\\ 1.6593530291522e-3 &  1.7197160649665e-3 &  1.47111490679553e-3 &  4.47958869966015e-3 \\
        \hline
        -3.58994878445356e-1 &  2.58807745278666e-3 &  1.32845068929825e-3 &  1.96278188235845e-3\\ 2.12047093908491e-3 & -3.59139073909566e-1 &  2.4482832432676e-3 &  1.92510275257336e-3\\ 2.03385597278184e-3 &  2.1239967070906e-3 & -3.59168172715776e-1 &  1.90480511326621e-3\\ 1.7197160649665e-3 &  1.77567805960787e-3 &  1.52879724701502e-3 & -3.59360882202121e-1 \\
        \hline
        1.12327371281389e-1 &  1.8085886792796e-3 &  1.07201615064092e-3 &  1.30812316112946e-3\\ 2.04441432655092e-3 & 1.12837363644479e-1 &  2.3553065012906e-3 &  1.63169569765752e-3\\ 1.72377197400074e-3 &  179094217772047e-3 & 1.12426473698445e-1 &  1.75603803032357e-3\\ 1.47111490679553e-3 &  1.52879724701502e-3 &  1.28467778185848e-3 & 1.12317122847217e-1 \\
        \hline
        4.44100160950618e-1 & -9.0800486359443e-2 & 4.81159771724384e-2 & 2.23151197796952e-1\\1.73175284635749e-1 & 3.12850716672893e-1 & 4.66282825250818e-1 &  9.92424706998542e-2\\1.58324230374422e-1 & 4.52211803714519e-1 & 3.83087886565564e-1 & -1.64191775784397e-1\\ 4.47958869966015e-3 & -3.59360882202121e-1 & 1.12317122847217e-1 & 7.9903188041306e-1 \\
        \hline
    \end{tabular} 
    \caption{Second half of a solution for N=4 Dirac fermions.}
    \label{table:4dirac2}
\end{sidewaystable}

\bibliographystyle{unsrt}

\end{document}